\documentclass[10pt,journal,compsoc]{IEEEtran}

\usepackage{cite}

\usepackage{amsfonts} 
\usepackage{graphicx} 

\usepackage{xcolor} 

\usepackage{titlepic}

\usepackage{stfloats} 

\usepackage{textcomp} 

\usepackage{multirow}

\usepackage{eso-pic}

\usepackage{wrapfig} 

\usepackage[ruled,vlined]{algorithm2e}

\usepackage{cite}

\usepackage{cryptocode}

\usepackage{titlepic}

\usepackage{multirow}

\usepackage{algpseudocode} 

\usepackage[utf8x]{inputenc}
\usepackage{enumitem}
\usepackage{amsmath}
\usepackage{graphics}

\usepackage{pifont}

\usepackage{tikz}
\usepackage{etoolbox}
\newcommand{\circled}[2][]{%
  \tikz[baseline=(char.base)]{%
    \node[shape = circle, draw, inner sep = 1pt]
    (char) {\phantom{\ifblank{#1}{#2}{#1}}};%
    \node at (char.center) {\makebox[0pt][c]{#2}};}}
\robustify{\circled}

\usepackage{tabularx,ragged2e,booktabs}
\newcolumntype{C}[1]{>{\Centering}m{#1}}

\usepackage{mathtools}

\usepackage{cryptocode}

\usepackage{url}
\usepackage{breakurl}
\usepackage[breaklinks]{hyperref}

\usepackage{xspace}
\usepackage{epstopdf}

\usepackage{comment}

\hfuzz=\maxdimen 
\newcommand{\todo}[1]{\textcolor{red}{TODO: #1}\PackageWarning{TODO:}{#1!}}

\newcommand{\damith}[1]{\textcolor{T}{Damith: #1}\PackageWarning{Damith:}{#1!}}
\definecolor{garrisonpink1}{rgb}{0.858, 0.188, 0.478}
\newcommand{\garrison}[1]{\textcolor{garrisonpink1}{Garrison: #1}\PackageWarning{garrison:}{#1!}}

\newcommand{\epc}{\textit{EPC Gen2}\xspace}

\definecolor{T}{RGB}{0,0,0}
\title{SecuCode: Intrinsic PUF Entangled \underline{Secu}re Wireless \underline{Code} Dissemination for Computational RFID Devices}

\author{Yang Su, Yansong~Gao, Michael Chesser, Omid~Kavehei, Alanson Sample and Damith C.~Ranasinghe

\IEEEcompsocitemizethanks{\IEEEcompsocthanksitem Yang Su is with Auto-ID Labs, School of Computer Science, The University of Adelaide, SA 5005, Australia. yang.su01@adelaide.edu.au}
\IEEEcompsocitemizethanks{\IEEEcompsocthanksitem Yansong~Gao is with School of Computer Science and Engineering, NanJing University of Science and Technology, Nanjing, China and Data61, CSIRO, Sydney, Australia.
yansong.gao@njust.edu.au}
\IEEEcompsocitemizethanks{\IEEEcompsocthanksitem Omid Kavehei is with School of Electrical and Information Engineering, The University of Sydney, NSW~2006, Australia. omid.kavehei@sydney.edu.au}
\IEEEcompsocitemizethanks{\IEEEcompsocthanksitem Alanson Sample is with Disney Research, Los Angeles, USA. alanson.sample@disneyresearch.com }
\IEEEcompsocitemizethanks{\IEEEcompsocthanksitem Damith~C.~Ranasinghe is with Auto-ID Labs, School of Computer Science, The University of Adelaide, SA 5005, Australia. damith.ranasinghe@adelaide.edu.au}
\IEEEcompsocitemizethanks{\IEEEcompsocthanksitem We acknowledge support from the Australian Research Council Discovery Program (DP140103448) and NJUST Research Start-Up Funding (AE89991/039)}\vspace{-0.5cm}
}

\begin{document}
\bstctlcite{IEEEexample:BSTcontrol}
\IEEEtitleabstractindextext{%
\begin{abstract}
The simplicity of deployment and perpetual operation of energy harvesting devices provides a compelling proposition for a new class of edge devices for the Internet of Things. In particular, Computational Radio Frequency Identification (CRFID) devices are an emerging class of battery free, computational, sensing enhanced devices that harvest all of their energy for operation. Despite wireless connectivity and powering, secure wireless firmware updates remains an open challenge for CRFID devices due to: intermittent powering, limited computational capabilities, and the absence of a supervisory operating system. We present, \textit{for the first time}, a \textit{\textbf{secure}} wireless code dissemination (SecuCode) mechanism for CRFIDs by entangling a \textit{device intrinsic hardware security primitive}—Static Random Access Memory Physical Unclonable Function (SRAM PUF)—to a firmware update protocol. The design of SecuCode: i) overcomes the resource-constrained and intermittently powered nature of the CRFID devices; ii) is fully compatible with existing communication protocols employed by CRFID devices—in particular, ISO-18000-6C protocol; and ii) is built upon a standard and industry compliant firmware compilation and update method realized by extending a recent framework for firmware updates provided by Texas Instruments. We build an end-to-end SecuCode implementation and conduct extensive experiments to demonstrate standards compliance, evaluate performance and security.

\end{abstract}


}

\maketitle
\IEEEdisplaynontitleabstractindextext

\section{Introduction}
\label{section:Introduction}
The exponential rate of hardware miniaturization, emergence of low cost and low power sensing modalities coupled with rapid developments in communication technologies are driving the world towards a future where {\it tiny scale computing} will be more pervasive and seamlessly integrated with everyday life. This sea of change is driven by the increasing ability of tiny computing platforms to connect people and things to the Internet---the Internet-of-Things (IoT)\cite{welbourne2009building}---enabling transformative applications ranging from healthcare \cite{islam2015internet,RanasinghePlosOneWearableSensor} to preventing counterfeiting \cite{yang2017cdta,jin2017secure}. Despite the magnitude of possibilities, the basic architecture of those IoT devices are quite similar; microcontrollers, transceivers, sensors, batteries, and sometimes, actuators are coupled with the most important component, the {\it application specific software} or more simply `code' which endows the `Things' with the ability to communicate with other parties and fulfill interactive tasks\cite{kortuem2010smart,zanella2014internet,lopez2011taxonomy}. 



More recently we have seen the emergence of such tiny scale computing platforms in the form of highly resource constrained and intermittently powered batteryless devices that rely only on harvested RF (radio frequency) energy for operations; best exemplified by  Computational Radio Frequency Identification (CRFID) devices such as WISP\cite{sample2008design}, MOO\cite{zhang2011moo} and commercial devices from Farsens\cite{farsens2016}. The benefit of batteryless devices arises from: i) the removal of expensive or risky maintenance, for example, when devices are deeply embedded in reinforced concrete structures or tasked with blood glucose monitoring or pacemaker control~\cite{tan2016wisent}; ii) the reduction in the cost of devices; and iii) the potential for an indefinite operational life. However, a significant challenge materializes from the need to patch, update or reprogram the application specific software in the form of firmware without the supervisory control of an operating system; to do so, a physical connection to a device is required. Unfortunately, a wired connection not only negates the benefit of the battery-free feature and makes the process unscalable when potentially millions of devices need to be updated, but is a more acute problem in the context of deeply embedded devices, such as a blood glucose monitor, where physical access to the device poses practical challenges and risks to the end-user. 



\subsection{Problem}
We define \textbf{firmware update} as the transfer of partial or entire executable code from the prover $\mathcal{P}$ to the non-volatile memory of the CRFID device, referred to as the token $\mathcal{T}$. The firmware update aims to, e.g., enhance CRFID device's computational or storage performance, impart new functionality, fix software bugs or address system compatibility. The firmware update is conducted either by \textit{physically connecting} a download cable or in a \textit{wireless} manner. Our work considers the more desirable but challenging secure wireless firmware update to eschew the cumbersome and, often impractical, cable-connected download method. 

\begin{figure}[h]
\centering
\vspace{-20pt}
\includegraphics[width=0.65\linewidth]{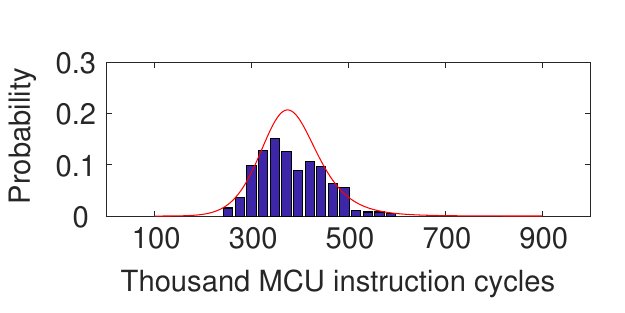}
\caption{\textcolor{T}{A distribution of clock cycles available for executing instructions on a CRFID device from power harvested during 100 different power-up and power loss events. We can see the limited and variable amount of energy or computational capability (measured as clock cycles) for executing code before losing state. (We used a WISP5-LRG \cite{parks2016} at 50~cm from a 9~dBic antenna energized by an Impinj R420 RFID Reader to collect the data, as described in~\cite{tan2016wisent}).}}

\label{fig:availableCycle}
\end{figure}
\begin{figure}[!h]
\centering
\includegraphics[width=0.8\linewidth]{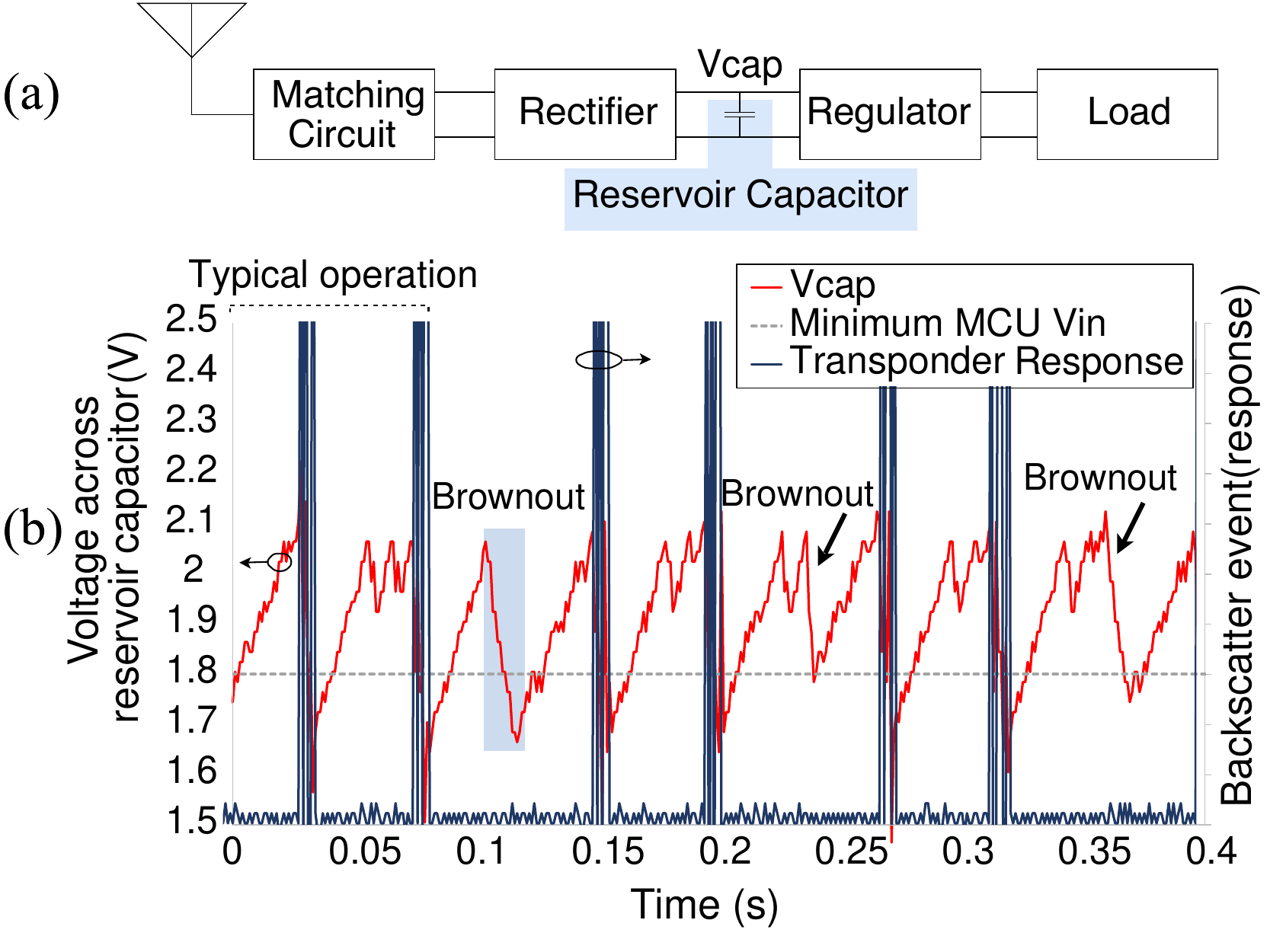}
\caption{(a) A typical power supply design based on harvested energy. (b) An example of sudden state loss as $V_{\rm{cap}}$ falls below 1.8~V on a CRFID transponder due to brownout events (the energy stored in the reservoir capacitor exhausted before the transponder completed its operation).}
\label{fig:suddenPowerLoss}
\end{figure}

The fundamental problem of a wireless firmware update for CRFID devices was addressed in two recent approaches~\cite{tan2016wisent,wu2017r}; both focused on transmission reliability, miniaturization of firmware code size, and energy efficiency. None of them addressed the difficult problem of assuring the security of wireless firmware updates, although Tan {\it et al.}~\cite{tan2016wisent} highlighted that secure wireless code dissemination remains the most urgent need to be addressed. This implies that both wireless approaches allow any party, irrespective of their trustworthiness, to remotely and wirelessly install code on a CRFID device. Consequently, malicious firmware injection from an adversary remains a direct threat that can lead to, for example, private information leakage such as health condition to unauthorized parties or the installation of malicious code in deeply embedded hardware such as a blood glucose monitor with devastating consequences for the victim. Therefore, this work aims to address the following questions: 

\begin{itemize}
\item {\it Is it possible to realize a secure and wireless code update mechanism without additional hardware components?}
\item {\it Can we develop an update protocol compliant with current communication protocols to ensure translation into practice?}
\item {\it Can we securely update firmware under resource constrains and intermittent powering, as illustrated in Fig. \ref{fig:availableCycle} and \ref{fig:suddenPowerLoss}, relying only on harvested power?}
\end{itemize}

Addressing the research questions above requires overcoming a number of challenges. We briefly outline the key challenges below:
\begin{itemize}
  \item Receiving and writing firmware to non-volatile memory (NVM) by a token requires considerable amount of energy\cite{wu2016r2} when operating on harvested power, even before we consider a  secure update mechanism. 
Battery-free CRFID transponders harvest energy from incident radio waves and the available energy is further influenced by a range of factors such as propagation path loss from the radio source to the transponder and shadowing effects created by intervening objects along the radio wave propagation path. Hence, devices that operate on harvested power, such as  CRFID transponders, are always constrained by energy and computation capability as illustrated in Fig.~\ref{fig:availableCycle}. Therefore, security must be realized under severe energy and computational resource limitations of a token. 
  \item Frequent state loss---as illustrated in Fig. \ref{fig:suddenPowerLoss}---results from interruptions to powering and the transient nature of power availability\cite{tan2016wisent}. This is especially problematic when updating executable code. Careless handling of state loss may lead to unrecoverable firmware corruption or leave the token vulnerable to attacks~\cite{akgun2013weaknesses}. Therefore, a firmware update mechanism should be robust even under state loss.
\item Security mechanisms rely on secure storage of keys. Traditional key storage methods in NVM is readily vulnerable to various attacks without expensive protection, for example, protective coatings and active tamper sensing circuity requiring batteries. Low-end IoT devices are usually lack of a secure NVM to maintain the key security. 
Therefore, a secure key storage method without additional hardware overhead, modifications and costs is highly desirable. 
\item A secure update protocol design must overcome the common difficulty of transferring large chunks of data to a transponder over narrow bandwidth wireless channels while still being compliant with existing standards and interface protocols for communicating with devices. Standards compliance will not only ensure successful adoption in practice but also interoperability with existing technologies. 

\end{itemize}


\subsection{Contributions}
Our work takes the first step towards secure code dissemination for resource constrained and intermittently powered CRFID devices through a systematic approach that combines theory and judicious cryptographic engineering. We summarize the contributions of our work below and defer comparison of prior work to Section~\ref{sec:relatedwork}.

\begin{itemize}
\item{\bf A first secure wireless firmware update protocol:} SecuCode is the first {\it secure} wireless firmware update or reprogramming method for CRFID devices to prevent malicious firmware injection attacks\footnote{\textcolor{T}{The security provided by the wireless firmware update scheme prevents an attacker from injecting malicious code. We achieve this aim by establishing the \textit{integrity} and \textit{authenticity} of a firmware. To reduce the burden on a resource limited device, we do not consider the provision of confidentiality protection.}}.
\item{\bf Lightweight physically obfuscated key derivation mechanism:} We develop a key derivation mechanism using a physical unclonable function (PUF). A PUF converts hardware instance specific random variations such as gate and wire delays of circuitry to a binary value. In particular, we take advantage of the random start-up state of intrinsic Static Random Access Memory (SRAM) on microcontrollers to extract key material on the fly without extra hardware overhead and modifications. The key derived from such an SRAM PUF is therefore: i) intrinsically tamper-resistant (unclonability) compared to a permanently stored key in non-volatile memory; ii) unique, through the hardware instance specific nature of SRAM cells' startup state; iii) unpredictable, through the physical randomness leading to random startup states of SRAM cells; and iv) never stored permanently in NVM as it is derived on the fly and discarded after usage. Our \textit{contribution} here is to address the challenges faced in realizing a lightweight, reliable and secure\footnote{\textcolor{T}{
Specifically, we refer to the security of the fuzzy extractor scheme expressed as the complexity of recovering the derived key by an attacker. Eventually, the security (authenticity and integrity) of the firmware update scheme rests on the security of the key derived from the reverse fuzzy extractor method.}} key generator using an on-chip SRAM PUF on a resource constrained device. 

\item{\bf Complete end-to-end design\footnote{Demonstration video:~\url{https://github.com/AdelaideAuto-IDLab/SecuCode/tree/master/demo}}:} We demonstrate an end-to-end design from firmware compilation to a successful firmware update process supported by an execution model on a CRFID token to manage the transient nature of power availability. We develop a tool (SecuCode App) to update firmware and conduct a complete end-to-end implementation and evaluation on a resource-constrained and intermittently powered CRFID device.
\item{\bf Standards compliance:} SecuCode is fully standards compliant: i) we implement SecuCode over the EPC C1G2v2 air interface protocol---ISO/IEC 18000-63:2015---commonly used by modern Radio Frequency Identification technology, including CRFID devices; and ii) given the specific Texas Instruments (TI) ultra-low power micro controller employed by CRFID devices, we implement our bootloader based on TI's recent framework for wireless firmware updates---\textit{MSP430FRBoot}~\cite{ryanbrownkatiepier2016}---to ensure a standard tool chain for compilation and update of new firmware whilst taking advantage of the features supported by MSP430FRBoot.
\item \textbf{Public code and SRAM PUF dataset release:}~We provide the complete end-to-end solution, including the SecuCode App source code, and research data collected on 21 devices to support future research in the field. The released material is available from~\cite{secucodegithub}.
\end{itemize}
\noindent{\bf Paper Organization:~}Section~\ref{sec:background} presents a brief background on CRFID system protocols and physical key derivation from a PUF. Section \ref{sec:seccode_protocol_design} presents the SecuCode protocol. Section \ref{sec:implementation} details the building  blocks required to realize SecuCode and their instantiation on a CRFID device. 
Section~\ref{sec:experiments} performs extensive experiments, including an end-to-end SecuCode implementation on a CRFID token as well as analyze its security. Section~\ref{sec:relatedwork} discusses related work. We conclude our work in Section~\ref{sec:conclusion}.



\section{Background}\label{sec:background}
\subsection{An Overview of CRFID System Protocols}\label{sec:c1g2}
\begin{figure}[t]
    \centering
    \includegraphics[trim={0.5cm 0 0 0},clip,width=1.04\linewidth]{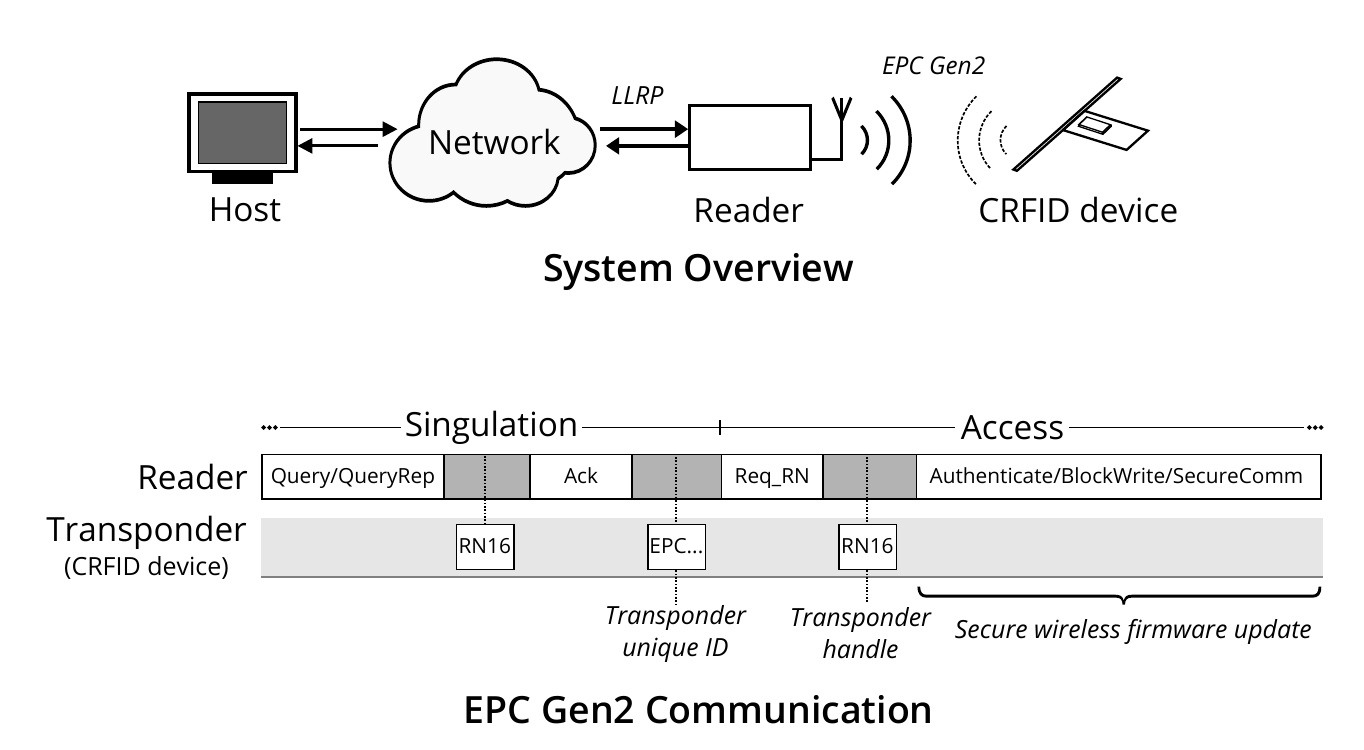}
    \caption{An overview of entities in the system (top) and an \epc protocol session (bottom).}
    \label{fig:system-overview}
\end{figure}


Fig.~\ref{fig:system-overview} illustrates a networked RFID system where the RFID reader resides at the edge of the network and connects to multiple antennas to communicate with long range RFID devices such as CRFID tags. The communication protocol between an RFID transponder operating in the UHF (Ultra High Frequency) range is governed by the widely adopted ISO-18000-6C---also known as the EPCglobal Class 1 Generation 2 version 2 (C1G2v2) air interface protocol or simply \epc protocol henceforth. While communication between a CRFID device and the reader takes place using the \epc protocol, the reader network interface described by the Low-Level Reader Protocol (LLRP) must be used by a host machine to communicate with a reader. Therefore, to realize a practicable solution, a secure code dissemination protocol must be implemented \textit{over} the \epc protocol where communication with readers must employ LLRP. Hence, a wireless firmware update process involves sending data from a host machine to a CRFID tag through a secure channel. This requires communication between three separate entities: i) the host  machine; ii) the reader; and iii)  the CRFID transponder.

The following steps are used to communicate with a CRFID device, or in general any \epc protocol compliant transponder:

\begin{itemize}
\item {\bf Host to Reader.} An application on the host machine constructs LLRP commands to build an \texttt{ROSpec} and \texttt{AccessSpec} to control the reader and transmits these specifications to a reader. 

\item {\bf Reader to CRFID.} As part of the anti-collision algorithm in the media access control layer, the reader must first singulate a CRFD device and obtain a handle, $RN16$. Singulation is achieved as part of the inventory operation to discover RFID devices in the powering and reading range of a reader. Inventory is performed through a combination of \texttt{Query}, \texttt{QueryRep}, and \texttt{QueryAdjust} commands. At the start of the inventory cycle the reader transmits a \texttt{Query} command, this notifies any devices in range, the beginning of a new inventory session. Each device then selects a random slot counter between zero and an upper value, $Q$, defined by the \texttt{Query} command. If the selected slot counter is zero, then the tag backscatters its handle, $RN16$, to the reader, otherwise the tag remains silent. On receiving a \texttt{QueryRep}, the tag decrements its slot counter and backscatters a response if the resulting counter value is zero. On receiving a \texttt{QueryAdjust}, the tag adjusts the $Q$ value and regenerates its slot counter.

\end{itemize}

A singulated transponder can subsequently be queried using a range of \textit{Access} commands such as \texttt{BlockWrite} used to write data to a transponder memory. Our approach is to realize a standard compliant secure firmware update protocol and therefore we design our secure firmware update protocol over the \epc protocol. Consequently, the firmware update process will occur after singulating the target transponder and employ \textit{Access} command specifications in \epc protocol.

\subsection{Physically Obfuscated Key Derivation}
In our secure wireless firmware update method, the integrity and authenticity of the firmware is dependent on the secure storage of a private key on a CRFID device. A physical unclonable function (PUF) is a cost-efficient security primitive for deriving {\it volatile} physically obfuscated keys on demand for resource and power limited devices. Therefore, we provide a brief overview on PUFs and introduce the reverse fuzzy extractor to realize secure and lightweight key derivation from  slightly noisy PUF responses (key material).

\subsubsection{Physical Unclonable Functions}
A PUF reacts with an instance-specific response (output) by exploiting manufacturing randomness when it is queried by a challenge (input)~\cite{suh2007physical}. It is widely employed for cryptographic key generation and lightweight authentication applications, including low-end resource-constrained devices~\cite{aysu2015end,van2012reverse,gao2017puf}. In this paper, we exploit the \textit{on-chip} SRAM to act as a PUF. Notably, the SRAM PUF is an \textit{intrinsic} PUF; i.e. the realization of the PUF does not require a custom design, additional chip area overhead, or hardware modifications~\cite{guajardo2007fpga,ruhrmair2012security}. Typically, one single SRAM cell consists of two cross-coupled inverters functioning as a latch. The initial power-up state of each SRAM cell is random but reproducible~\cite{holcomb2007initial}, thus, the power-up pattern of bits generated from an SRAM memory---referred as an SRAM PUF---can be viewed as a unique identifier. In such a PUF, the address of the SRAM cell acts as the challenge, the initial power-up state---`1'/`0'---acts as the response. \textcolor{T}{It is worth mentioning here that, instead of the power-up state of the SRAM cells, the data retention voltage (DRV) of SRAM cells can also be characterized and exploited as PUF responses---the so-called DRV PUF~\cite{holcomb2012drv,xu2015reliable}. In our study, we employ the power-up state of SRAM to build a PUF due to the simplicity of the response readout method at power-up}. However, use of SRAM PUF responses as key material is challenged by inconsistent regeneration of response bits and potential response bias, which are addressed in Section~\ref{sec:lightweightAndSecurePOK}.
\subsubsection{Reverse Fuzzy Extractor}\label{sec:reversefuzzyex}
PUF response regeneration is noisy as it is vulnerable to fluctuations in environmental conditions such as thermal noise and power supply and temperature variations. Thus, PUF responses cannot be directly used as a key and  requires error correction to rectify flipped response bits (errors) in relation to a reference response. In this case, a fuzzy extractor (FE) is usually deployed to derive a key by using helper data to correct regenerated responses at a later time.

A fuzzy extractor \textsf{FE} is described by two algorithms---see detailed descriptions in Section~\ref{sec:concepts_notations}: i) key generation or enrollment algorithm \textsf{FE.Gen}; and  ii) key reconstruction algorithm \textsf{FE.Rec}. 
However, the computational complexity of computing the helper data by \textsf{ FE.Gen} and the key reconstruction by \textsf{ FE.Rec} using the helper data is asymmetric. Therefore, here we construct a reverse fuzzy extractor~\cite{van2012reverse} to ensure the lightweight \textsf{ FE.Gen} implementation is on the resource-constrained CRFID device while the computationally expensive \textsf{FE.Rec} is implemented at the resource-rich server (or prover $\mathcal{P}$ in our case).

\textcolor{T}{In general, a reverse fuzzy extractor's \textsf{FE.Gen} function is implemented on the device. Then the device applies a random challenge \textbf{c} sent by a server to obtain a noisy response ${\bf r}$ from an on-device PUF and uses the generator function \textsf{FE.Gen} to compute helper data \textbf{h} for the noisy response ${\bf r}$. The helper data \textbf{h} is sent to the server. Subsequently, the server uses the helper data to reconstruct the response ${\bf r}$ with a securely stored response ${\bf r}^{\prime}$ that is evaluated during a secure provisioning phase for the given challenge \textbf{c}~\cite{van2012reverse}.}

\section{SecuCode: Protocol Design}
\label{sec:seccode_protocol_design}
We ease into the design of the protocol by first describing the notations used in the protocol followed by a formal description of the SecuCode protocol.

\subsection{Notations}\label{sec:concepts_notations}
The notations used in the rest of the paper, following that in~\cite{aysu2015end}, is described below:
\newline
\begin{itemize}
\item A bold lowercase character is used for a vector, for example, a challenge applied to a PUF is {\bf c}. A bold uppercase character is used for a set, for example, a challenge set {\bf C}, where {\bf c~}$\in${\bf ~C}.

\item \textbf{True Random Number Generator} \textsf{TRNG}
outputs a truly random number when invoked.

\item \textbf{Physical Unclonable Function} \textsf{PUF} takes a challenge {\bf c} as input and reacts with a corresponding response {\bf r} as output, where {\bf r} $\leftarrow$ \textsf{PUF}({\bf c}).



\item \textcolor{T}{ \textbf{Message Authentication Code} \textsf{MAC} computes a bit string $s$ of fixed length to establish the authenticity and the integrity of a message. Specifically,  $s\leftarrow${\textsf{MAC}({\bf m}, {\bf sk})} computes a MAC from a message ${\bf m}$---to prevent replay attacks, a nonce can be concatenated with the message---and a key ${\bf sk}$.}


\item \textbf{Fuzzy Extractor} A fuzzy extractor \textsf{FE} is defined by two functions: key generation algorithm \textsf{FE.Gen} and key reconstruction algorithm \textsf{FE.Rec}. The \textsf{FE.Gen} takes a variable ${\bf z}$ as input and produces key ${\bf sk}$ and helper data ${\bf h}$. The \textsf{FE.Rec} algorithm recovers the $\bf sk$ assisted with ${\bf h}$ by taking the ${\bf z}'$ as input if the Hamming distance \textsf{HD}(${\bf z},{\bf z}'$) is sufficiently small, i.e. \textsf{HD}(${\bf z},{\bf z}'$)$\leq~t$ with $t$ a fixed parameter---\textit{correctness of reconstruction guarantee}. If \textsf{HD}(${\bf z},{\bf z}'$)$\leq~t$	for input {\bf z} and min-entropy~\cite{delvaux2016efficient} of ${\bf z}\geq|{\bf h}|$, the fuzzy extractor provides key $\bf {\bf sk}$ that is statistically close to a uniformly distributed random variable in $\{0,1\}^{|\textbf{sk}|}$ although helper data $\textbf{h}$ is exposed to an adversary---\textit{security guarantee}. Fuzzy extractors are generally built with an error-correction method~\cite{delvaux2015helper}.
\end{itemize}

\subsection{Adversary Model}\label{sec:adversarymodel}

In this paper, our focus is on the communication between the Reader and the CRFID transponder or Token $\mathcal{T}$. We assume that the communication between a Host and a Reader is secure using standard cryptographic mechanisms for securing communication between two parties over a network~\cite{jin2017secure}; hence a Host computer and a Reader are considered as a single entity, the Prover, denoted as $\mathcal{P}$. 

There is no previous CRFID firmware update protocol that considers security. Therefore, no existing adversary model has been reasoned. In this initial secure wireless firmware update investigation, we follow a relevant model and assumptions in PUF-based authentication protocols designed for resource-constrained platforms~\cite{van2012reverse,aysu2015end}.

Notably, a wireless firmware update of a CRFID device is only possible after: i) the commissioning of the device whereby an immutable program called the \textbf{bootloader} is installed on the device; and ii) the prover $\mathcal{P}$ has enrolled---extraction and secure storage of---SRAM PUF responses in a secure environment 
using a one-time access wired interface. We assume that the wired interface is disabled after the installation of the bootloader and enrollment of the PUF responses. In other words, the adversary $\mathcal{A}$ cannot directly access the SRAM PUF responses, only the immutable bootloader maintains this access at power-up and for a very short duration of time. After the commissioning of the device, both a trusted party and the adversary $\mathcal{A}$ must use the wireless interface for installing new firmware on a token.

Subsequent deployment of a token $\mathcal{T}$ will place it in an adversarial environment where only the prover $\mathcal{P}$ remains trusted. We assume that the attacker $\mathcal{A}$ can eavesdrop on the communication channel, isolate the CRFID transponder from the system and carry out a man-in-the-middle attack and forward  tampered information from $\mathcal{P}$ to $\mathcal{T}$ and vice versa. Further, following the assumptions in~\cite{aysu2015end}, within the adversarial environment, the adversary  $\mathcal{A}$ may obtain any data stored in the NVM of the devices. However, as in~\cite{aysu2015end}, the adversary $\mathcal{A}$ cannot mount implementation attacks against the CRFID, nor gain internal variables stored in the registers, for example, using invasive attacks and side-channel analysis. \textcolor{T}{Similar to other adversarial models, we do not consider Denial of Service (DoS) attacks because, in practice, it is not possible to defend against an attacker that, for example, disrupts or jams the wireless communication medium~\cite{kohnhauser2016secure}, or attempts to heat the intrinsic SRAM cells to prevent a firmware update as a result of the proposed conditional firmware update method (detailed in Section~\ref{sec:preselection}).}



\subsection{SecuCode Protocol}
\label{sec:seccode_protocol}

\begin{figure}[!ht]
    \centering
    \includegraphics[width=0.98\linewidth]{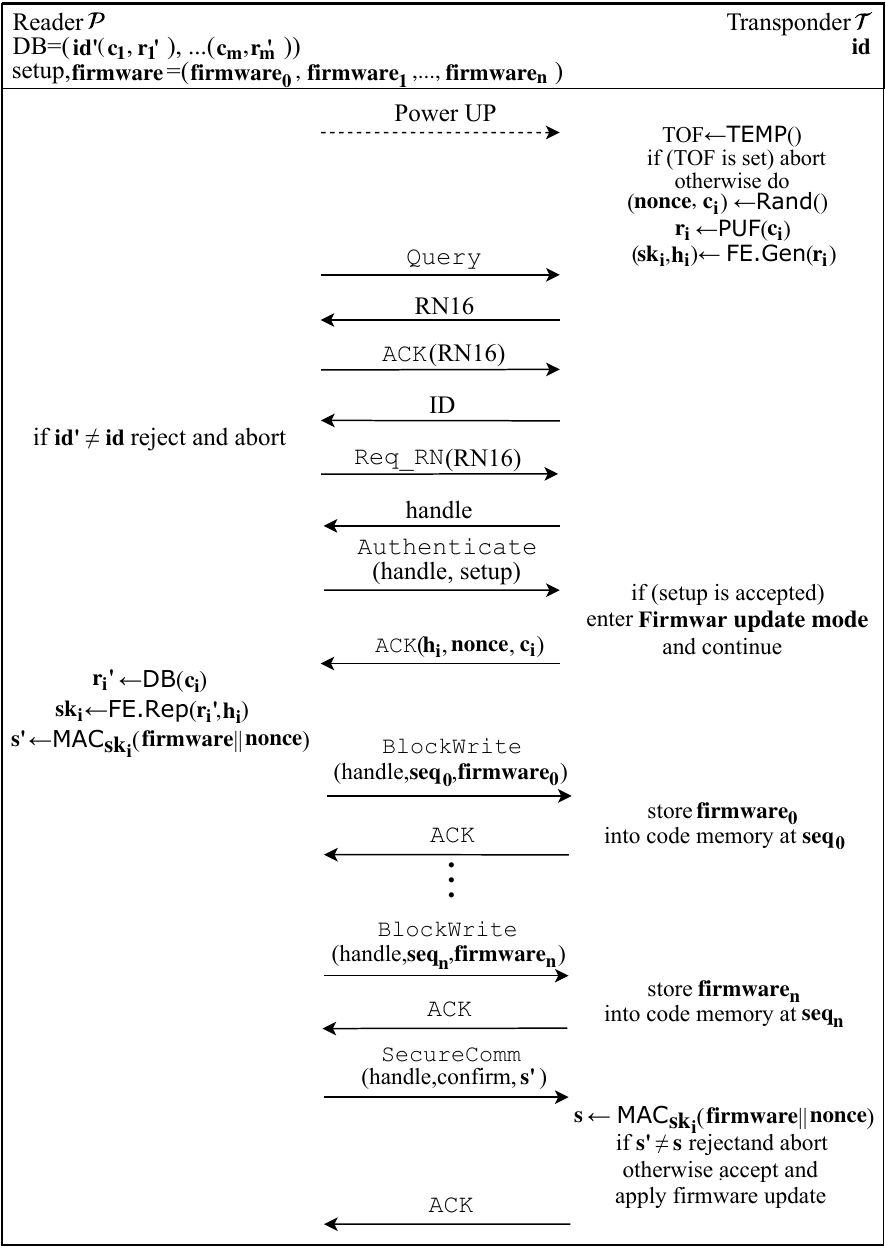}
	\caption{SecuCode protocol. }
    \label{fig:IPno}
\end{figure}

\textcolor{T}{SecuCode protocol described in~Fig.~\ref{fig:IPno} relies on the simplicity of transmitting the firmware in plaintext and assumes the adversary $\mathcal{A}$ can gain full knowledge of the firmware---this is consistent with our adversary model which assumes that an adversary can read the contents of the NVM of a CRFID device. Our focus is to prevent malicious code injection attacks. SecuCode achieves this goal by facilitating the authentication of the prover by the token and ensuring the integrity of the firmware by the token before accepting the firmware. Therefore, only firmware issued by the trusted prover will be accepted.} 

The SecuCode protocol is designed to be implemented over the \epc protocol. We employ the recently defined extended \textit{Access} command features for supporting future security services on RFID transponders. Consequently, firmware update initializations employs \texttt{TagPrivilege} and \texttt{Authenticate} while downstream data transmissions in SecuCode are carried out by employing the \epc protocol specification of \texttt{BlockWrite} and \texttt{SecureComm} commands. \textcolor{T}{The \texttt{SecureComm} command specification allows the encapsulation of other \epc protocol commands, such as \texttt{BlockWrite} but the payload is encrypted. Hence, we employ the \texttt{SecureComm} command to transport the message authentication code to the token.} 
We employ \texttt{BlockWrite} command to write firmware to a memory space, or download area, allocated and managed by the bootloader on a CRFID transponder. Although the specification of \texttt{TagPrivilege}, \texttt{Authenticate} and \texttt{SecureComm} commands are defined in the \epc protocol, it is important to mention here that these commands {\it are yet to be supported} on CRFID transponders and this also constitutes one of the tasks in this study.
The key phases in our proposed SecuCode protocol are summarized below: 


\begin{itemize}
\item \textbf{Prover initialization phase in a secure environment}: 
This phase is carried out in a secure environment. A publicly known unique ID string is stored in a token's NVM. The prover $\mathcal{P}$ enrolls in a database DB the ID string of the target token $\mathcal{T}$ as well as the challenge-response pairs (CRP)  from the PUF (also known as the enrollment phase~\cite{suh2007physical}). The bootloader, immutable program stored in a write protected memory space, is installed on the token $\mathcal{T}$ by the prover $\mathcal{P}$  and, subsequently, the physical interface to  $\mathcal{T}$ is disabled. The bootloader is responsible for the SecuCode protocol implementation on the token.

\item \textbf{Firmware update phase in a potential adversarial environment}: For each code dissemination session, there will be a compiled firmware at the prover $\mathcal{P}$ to be transmitted along with a setup profile which describes the size of the firmware, starting memory address and the MAC method for the token $\mathcal{T}$. In particular, the following occurs: i) \textbf{lightweight physically obfuscated key derivation} on the token and the subsequent transmission of the token generated random challenge seed $\bf c$ and helper data $\bf h$ to the prover; ii)~\textbf{firmware update} which includes the wireless transfer of firmware to the token, the establishment of the veracity of the firmware on the token to accept/abort the firmware update issued by the prover and update of firmware on the token. We elaborate on these stages below.  
\end{itemize}

\subsubsection{Lightweight Physically Obfuscated Key Derivation}
After the token $\mathcal{T}$ harvests adequate power from the prover $\mathcal{P}$, a $\bf nonce$ is generated for use in the firmware update session, meanwhile a random number \boldmath$\bf{c_i}$ is generated as the seed challenge; $\bf c_i$ can be viewed as a challenge seed that determines the starting index into a byte level address in a block of highly reliable and unbiased SRAM PUF cells; for a detailed discussion please refer to Section~\ref{sec:lightweightAndSecurePOK}. These responses are subsequently readout; ${\bf r_i} \leftarrow $\textsf{PUF}($\bf c_i$).
 
We propose a \textbf{\textit{conditional firmware update}} method based on evaluating the on-chip temperature prior to the key derivation phase since response reliability of SRAM PUFs are more sensitive to changes in temperature than supply voltage. This is to significantly reduce the computational burden on the PUF key generation overhead while meeting error correction and security bounds---details in Section~\ref{sec:lightweightAndSecurePOK}. Immediately before key derivation, the $\textsf{TEMP()}$ function sets an over-temperature flag (TOF) if the in-built chip thermometer reports a temperature outside of a legal range ( $0^{\circ}$C to $40^{\circ}$C is the chosen legal range in this work). Setting OTF will result in aborting key derivation and triggering a re-booting of the token. 

A token $\mathcal{T}$ operating under a legal temperature range will execute the private key $\bf sk_i$ derivation and helper data $bf h_i$ generation as ($\bf sk_i$,$\bf h_i$)$ \leftarrow $\textsf{FE.Gen}($\bf r_i$). The private key $\bf sk_i$ is used in the following firmware update process and only retained in the SRAM on the token $\mathcal{T}$ for the duration of the protocol session and discarded: i) at the completion of a session; or ii) during a power loss event.



\subsubsection{Firmware Update}
The ID of the token $\mathcal{T}$ is checked by the prover $\mathcal{P}$ to select the target token and once the target is confirmed to be visible to the prover and is singulated, the prover $\mathcal{P}$ can employ \textit{Access} commands \texttt{Authenticate}, \texttt{BlockWrite} and \texttt{SecureComm}---see Figure~\ref{fig:system-overview} for an illustration of the \epc protocol behavior---to execute the firmware update. We describe the update phase below.

The prover issues an \texttt{Authenticate} command to deliver setup parameters. The token responds with the $nonce$, $\bf c_i$ and $\bf h_i$ back to the prover. 
The prover $\mathcal{P}$ reconstructs the private key $\bf sk_i$ through $\bf sk_i\leftarrow$ \textsf{FE.Rec}($\bf r_i'$, $\bf h_i$); here, $\bf r_i'$ is the enrolled response corresponding to $\bf c_i$. Now the prover $\mathcal{P}$ and the token $\mathcal{T}$ have a shared private key.

The firmware cannot generally be sent to the token $\mathcal{T}$ in a single transaction. The \epc protocol implies a limitation on the length of a payload string to 255 words; this can be inadequate to encapsulate a practical CRFID firmware\cite{tan2016wisent}. Therefore, we partition the firmware input into $n$ chunks \{$\bf firmware_0$, $\bf firmware_1$, $\bf firmware_2$, ..., $\bf firmware_n$\} and transmit sequentially indexed chunks $\{seq_0,seq_1,...,seq_n\}$ to the token $\mathcal{T}$ using the \texttt{BlockWrite} command. Here, $seq_i$ indicates the relative offset of firmware chunk $\bf firmware_i$. 

Before the firmware update is applied, the token $\mathcal{T}$ must validate the authenticity of the prover $\mathcal{P}$ and the integrity of the firmware. \textcolor{T}{First, a message authentication code $s^\prime$ is computed by the prover $\mathcal{P}$ using a \textsf{MAC} function as ${\bf s'}\leftarrow$\textsf{MAC}($\bf firmware\|nonce$, $\bf sk_i$), with $\bf sk_i$ the reconstructed PUF key. Second, the prover sends $s'$ to the token. The token computes ${\bf s}=$\textsf{ MAC}($firmware\|nonce$, ${\bf sk}_i$) locally to compare $\bf s$ and $\bf s'$. If $\bf s$ and $\bf s'$ match, the token $\mathcal{T}$ accepts and applies the firmware update.} Success of a firmware update is signaled to the prover $\mathcal{P}$ by the token backscattering an \texttt{ACK}. This process ensures:

\begin{enumerate}
\item The integrity of the received firmware at the token $\mathcal{T}$. Any corruption or mutation will lead to the failure of the integrity check on the token owing to the \textsf{MAC} and subsequent discarding of the firmware.
\item The authority of the prover $\mathcal{P}$. Only the trusted prover can obtain same secret key $\bf sk_i$ to issue a valid MAC $\bf s'$. 

\end{enumerate}

\section{Implementation}
\label{sec:implementation} 
In this section: 
\begin{enumerate}
\item We generalize the SecuCode control-flow on a token $\mathcal T$ and provide an overview of the required functional blocks. 
\item We describe the challenges in instantiating the functional blocks and propose approaches to address them.
\item We complete the end-to-end implementation based on the instantiated functional blocks
\end{enumerate}

We have selected the open-hardware and software implementation of WISP5.1-LRG~\cite{parks2016} CRFID transponder as our token $\mathcal{T}$ for a concrete implementation and experiments. This intermittently powered CRFID transponder is built using the ultra-low power microcontroller unit (MCU) MSP430FR5969 from Texas Instruments. Therefore, when required, we provide specific implementation examples on a CRFID transponder based on the WISP5.1-LRG and MSP430FR5969 MCU in the discussions that follow.
\subsection{Protocol Control-Flow on a Transponder}
\label{sec:control-flow} 

\begin{figure}[!ht]
    \centering
    \includegraphics[width=\linewidth]{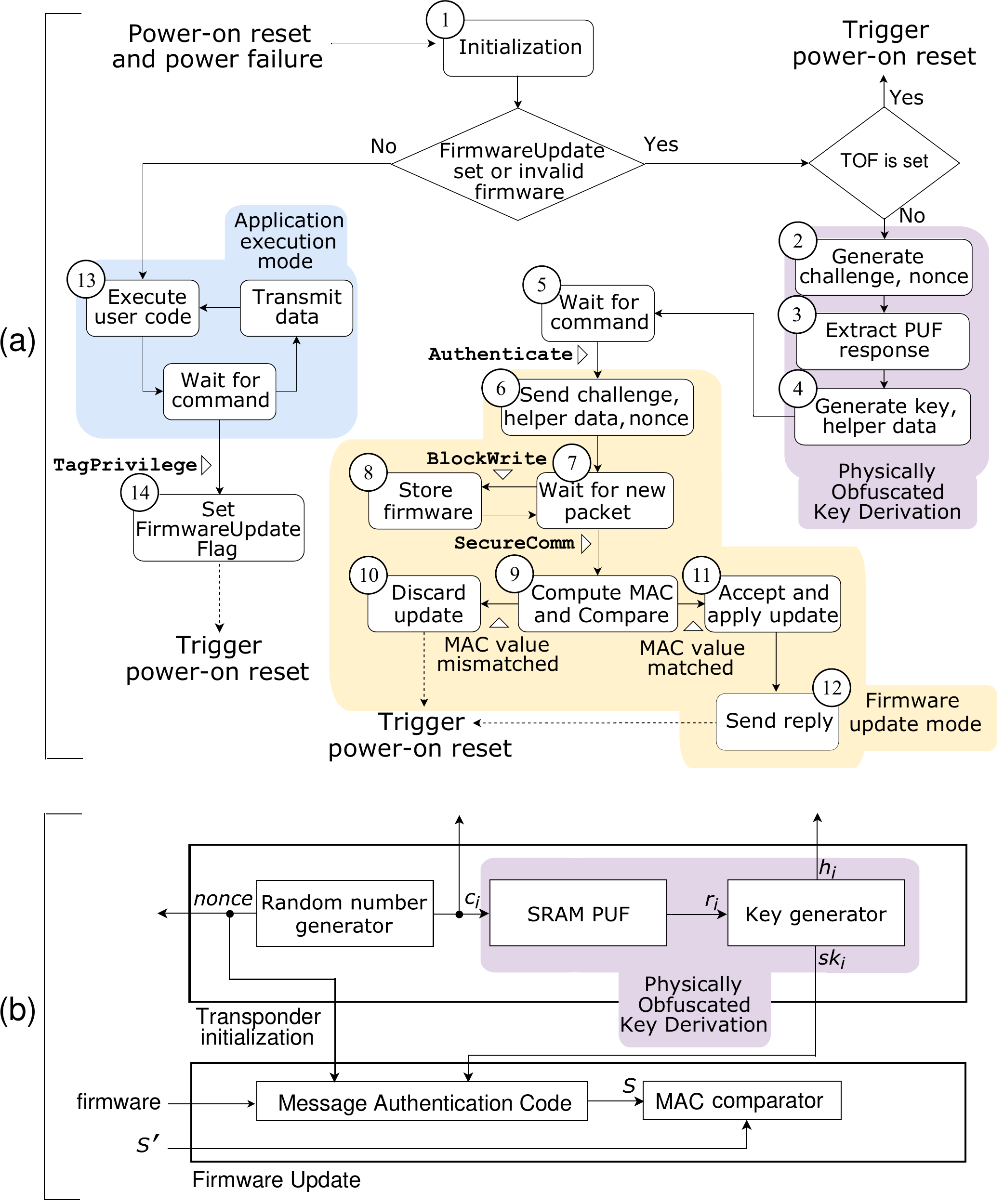}
    \caption{(a) Control flow and (b) Protocol functional blocks implemented to instantiate SecuCode protocol on a CRFID transponder. }
    \label{fig:Flowchart}
\end{figure}

Following the SecuCode protocol in Section~\ref{sec:seccode_protocol}, the generalized control flow on a transponder is illustrated in Fig. \ref{fig:Flowchart} and detailed below. 

\circled{1} When a token $\mathcal{T}$ is adequately powered, it initializes the MCU hardware, and determines whether to run in firmware update mode or application execution mode. If TOF is not set (the token is operating in a legal temperature range), the token enters the key derivation phase and \circled{2} a challenge $\bf c_i$ and a $nonce$ are generated using the \textsf{TRNG}. \circled{3} The token $\mathcal{T}$ reads the response $\bf r_i$ corresponding to $\bf c_i$. \circled{4} The token $\mathcal{T}$ derives a key $\bf sk_i$ and computes helper data $\bf h_i$ through ($\bf sk_i,\bf h_i$)$\leftarrow$\textsf{FE.Gen}($\bf r_i$). These steps complete the key derivation phase and \circled{5} the token $\mathcal{T}$ awaits for further commands from the prover $\mathcal{P}$.

 \circled{6} The token $\mathcal{T}$ responds to an \texttt{Authenticate} command consisting of the firmware update setup parameters with $\bf h_i$, $\bf c_i$, and $nonce$ to enable the prover $\mathcal{P}$ to reconstruct the PUF key $\bf sk_i$. The \texttt{Authenticate} command also directs the token  $\mathcal{T}$ to enter the \textbf{Firmware Update} mode. \circled{7} Whilst in this state, the prover $\mathcal{P}$ can transmit new firmware in chunks. 
 \circled{8} Given a firmware chunk, $firmware_i$, it is transmitted using \texttt{BlockWrite} commands and stored in the download area in a receive and store process. \textcolor{T}{At the completion of the wireless firmware transmission, the MAC value $s'$ computed by the prover is encapsulated in a \texttt{SecureComm} command and sent to the token~$\mathcal{T}$}. 

\circled{9} The MAC value $s$ computed by the token using the received $firmware$ and $nonce$ with the secret key $\bf sk_i$ is compared; if the integrity of the firmware and authenticity of the prover is established, \circled{11} the $firmware$ is accepted and \circled{12} the token sends an \texttt{ACK} to the prover. Otherwise, \circled{10} the update is discarded. Regardless of the acceptance or rejection decision, the token exits the {\bf Firmware Update} mode by triggering a power-on-reset. 


\circled{13} If a firmware update is not required,  the token $\mathcal{T}$ executes the user code, \circled{14} if the token receives a \texttt{TagPrivilege} command, indicating entry into firmware update mode, the token restarts in firmware update mode by setting a firmware update flag---\textit{FirmwareUpdate}---and triggering a power-on-reset. Whenever the token $\mathcal{T}$ is rebooted by a \texttt{TagPrivilege} command to enter the \textbf{Firmware Update} mode, $nonce$ and derived key $\bf sk_i$ is refreshed. The $\bf sk_i$ changes because: i) the challenge seed is refreshed; and ii) a varying response is produced even for the same challenge as a consequence of the naturally noisy nature of the response bits.

\textcolor{T}{A power failure such as a brownout event, as shown in Fig.~\ref{fig:suddenPowerLoss}, during the execution of the protocol will result in a reset and rebooting of the token $\mathcal{T}$. In such an event, the immutable bootloader's functionality is preserved. Therefore, a prover $\mathcal{P}$ can attempt another secure firmware update.} In the following sections, we describe the instantiation of each functional block shown in Fig.~\ref{fig:Flowchart}(b).

\subsection{Random Number Generator}\label{sec:SRAMTRNG}
\textcolor{T}{Our implementation of the SecuCode protocol shown in Fig.~\ref{fig:IPno} employs an 8-bit challenge seed and a 128-bit $nonce$.} As with other low-end computing platforms, acquiring true random numbers on a CRFID device is challenging given the lack of resources to implement a cryptographically secure random number generator (RNG). Ideally, the RNG should be a true random number generator (TRNG) and its implementation should require no modifications to existing hardware. We evaluate and summarize the performance of three RNGs in Table~\ref{tab:RNG} in terms of random bits per request, time overhead, power consumption and required hardware modules in Appendix~\ref{Sec:APX-TRNG}.



We can see that the SRAM TRNG, implementation based on the study in~\cite{aysu2015end}, outperforms the rest with regards to time and energy overhead. Most importantly, it requires \textit{no extra hardware}. Therefore, an SRAM TRNG is chosen for implementing the SecuCode protocol.

\subsection{Lightweight Physically Obfuscated Key Derivation}
\label{sec:lightweightAndSecurePOK}
Deriving and sharing a private key between the prover $\mathcal{P}$ and the token $\mathcal{T}$ should also be: i) lightweight; and ii) secure. 
Realizing both requirements on a resource limited token is challenging. Thus we: i) employ an SRAM PUF to derive a key instead of a stored key in non-volatile memory---prone to extraction through physical means and requiring expensive secure NVM---with the ability to refresh the key between protocol sessions; and ii) employ a reverse fuzzy extractor to realize a lightweight \textsf{FE.Gen} on the token to compute helper data necessary for the server to independently reconstruct the shared key with very high probability of success. 
In particular, we propose the following mechanisms to derive a lightweight and secure PUF key:

\begin{itemize}
\item \textbf{Enhancing reliability of key material:} Time complexity of the generator function \textsf{FE.Gen} and  security---information leakage---is related to the amount of helper data needed to correct noisy SRAM PUF response bits. Therefore, we winnow SRAM PUF responses with high bit specific reliability using \textit{response pre-selection with an on-chip selection meta-data storage structure} together with our \textit{conditional firmware update} method to significantly reduce the demand on helper data (see Section~\ref{sec:preselection}). 
\item \textbf{Removing information leakage through response bias:} PUF response bias, an imbalance between the number of zeros and ones, has shown to leak additional information~\cite{delvaux2016efficient,delvaux2015helper}. Therefore, to guarantee the security bounds of the reverse fuzzy extractor, we  propose a Hamming weight based response de-biasing method to eliminate response bias (see Section~\ref{sec:HWde-biasing}).
\end{itemize}


\subsubsection{Enhancing Reliability}
\label{sec:preselection}
We consider a syndrome based construction as in~\cite{van2012reverse} and use a BCH($n,k,t$) linear block code encoder to build \textsf{FE.Gen}. The \textsf{FE.Gen} function is responsible for generating the helper data $\bf h$ used by the prover to reconstruct the PUF response extracted by the token using a previously enrolled response that is securely stored at the prover. 
Here, $t$ denotes the number of errors a BCH($n,k,t$) code is capable of correcting, 
$n$ denotes the number of bits extracted from a PUF or the length of the response $\textbf r_i$ where the length of the helper data is $ |{\bf h}|= (n - k)$--- which also defines the well known upper bound on information leakage. 

Although we can select a BCH($n,k,t$) code with appropriately large parameter values to achieve the desired attack complexity, error correcting capability to achieve an industry standard key failure rate of less than $10^{-6}$ and number of key bits $k$, the computational time complexity of a BCH encoder, $\mathcal{O}(n^2)$, forces the use of parallel blocks of BCH($n,k,t$) code with smaller values of $n$. For $|{\bf r_i}|/n$ parallel blocks of BCH($n,k,t$) code using a syndrome construction, the complexity of finding ${\bf r_i}$ is $2^{k|{\bf r_i}|/n}$~\cite{delvaux2015helper}.

The key failure rate when employing a BCH($n,k,t$) code is expressed as:
\begin{equation}
	P_{1} = 1-{\rm binocdf}(t,n,{\rm BER})
\end{equation}
where {\tt binocdf} is the binomial cumulative distribution function, \textcolor{T}{BER (bit error rate) is the response unreliability expressed as~\cite{gao2015memristive}:
\begin{equation}
	{\rm BER}=\frac{1}{m}\sum_{t=1}^{m}\frac{{\mathsf{HD}}({\bf r},{\bf r}'_{t})}{|{\bf r}|}
    \label{eqn:ber}
\end{equation}
where $m$ is the number of response evaluations under a given operating corner (a combination of voltage and temperature), ${\bf r}$ is the reference response (typically the nominal operating voltage and temperature), and ${\bf r'_t}$ is the $t^{\rm th}$ regenerated response under different operating corner.} 
As we use multiple blocks, the failure rate when $|{\bf r_i}|/n$ blocks are employed is expressed as:
\begin{equation}
	P_{\rm Fail} = 1-(1-P_{1})^{|{\bf r_i}|/n}
    \label{eq:p-fail-for-parallel}
\end{equation}

Therefore, we can see that it is imperative to reduce BER to significantly decrease the key failure rate $P_{\rm Fail}$ and reduce the complexity of the BCH encoder required. 
We devise the following \textbf{two} methods to significantly reduce the complexity of the BCH encoder required on the resource constrained token: i) response pre-selection together with on-chip selection meta-data storage structure; and ii) a conditional firmware update strategy. As shown by detailed experimental evaluations in Section~\ref{sec:ExReliability}, our approaches significantly reduce the expected BER to be $<1\%$.

\vspace{0.2cm}
\noindent\textbf{Pre-selection:} SRAM PUF pre-selection was first noted by Hofer {\it et al.} in~\cite{hofer2010alternative}. The idea is to locate SRAM cells which tend to generate stable PUF responses. During the enrollment phase, unstable responses are identified and discarded for key generation\cite{bohm2013puf}. 

\begin{figure*}[!t]
    \centering
    \includegraphics[width=.8\linewidth]{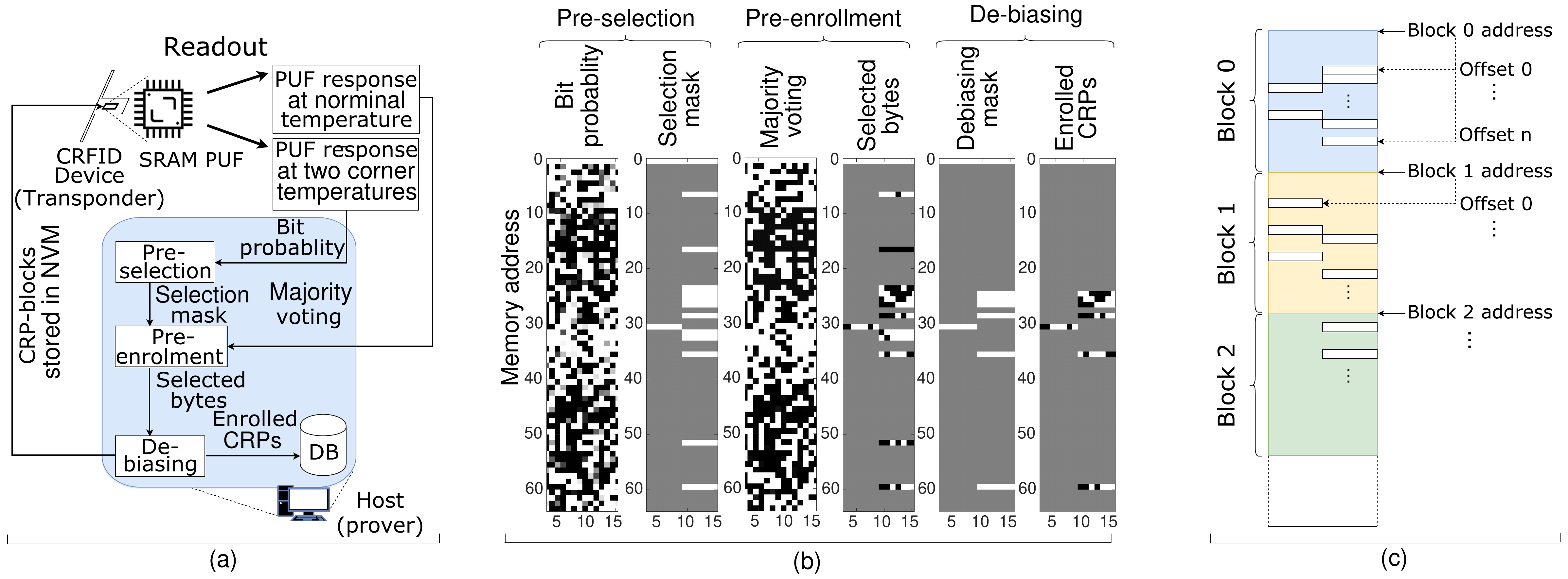}
    \caption{(a) Procedure for pre-selection, enrollment and de-biasing to enhance the response reliability and generate unbiased PUF response. (b) Memory maps detailing each step---only the first 64 words are shown here. (c) CRP-block map, a compact data structure to store the mask configuration on the CRFID transponder for fast response readout.}
    \label{fig:preSelection}
\end{figure*}

We employ an approach similar to the multiple-readout method in~\cite{bohm2013puf}. Given that a microcontroller's SRAM memory is byte addressable, we employ a byte-level response selection method illustrated in Fig.~\ref{fig:preSelection}. In particular, we first select response bytes that are reliably reproduced under repeated measurements 
generated under two corner temperatures---selected as 40°C and 0°C in our implementation. Subsequently, under nominal temperature (25°C), the former response bytes are further subjected to multiple readouts and majority voting (e.g., if 8 out of 10 readouts of a bit yields logic `1', then this bit is enrolled as a logic `1') is applied enrol values for the response bytes.

\vspace{0.2cm}
\noindent\textbf{Conditional firmware update:} It is recognized that the BER of SRAM PUF is sensitive to temperature but insensitive to supply voltage variations~\cite{cortez2012modeling}. Therefore, in addition to reliable response pre-selection, we propose performing a conditional firmware update based on the core operating temperature of the token. A firmware update is executed only when the core temperature of the chip is within a legal temperature range; considering most practical applications, we selected 0°C to 40°C range for SecuCode. Without extra hardware overhead, we take advantage of the available temperature sensor within the MSP430FR5969 microcontroller used by the CRFID transponder to sample the temperature before the PUF responses are readout. \textcolor{T}{A temperature breaching the legal range terminates further execution and triggers a power-on-reset.} 


\subsubsection{De-biasing}\label{sec:HWde-biasing}
Most studies assume that PUF responses are uniformly distributed and hence an n-bit response has the fully entropy of $n$ bits. Using a BCH($n,k,t$) code, ($n-k$)-bit helper data ${\bf h}$ will be publicly known. Under the disclosure of helper data ${\bf h}$, there is no less than k-bit entropy remaining. However, as highlighted recently~\cite{maes2015secure,delvaux2016efficient}, response bias incurs extra entropy loss. Thus, the $k$-bit min entropy bound guaranteed by a BCH($n,k,t$) encoder might be decreased. To prevent extra entropy leakage, response de-biasing methods can be employed~\cite{maes2015secure,aysu2017new}. In general, de-biasing converts a response $\bf r_x$ into an unbiased enrolled response $\bf r_y$, where response $\bf r_x$ might have a bias and $|\bf r_x|\geq |\bf r_y|$. De-biasing needs to consider four aspects: i)~{\it de-biasing should not deteriorate or increase response error rate}; ii) {\it efficiency}; iii) {\it information leakage}; and iv) {\it reusability}~\cite{maes2015secure,delvaux2016efficient}; to this end, we propose the following Hamming weight (HW) based de-biasing method.

\vspace{0.3cm}
\noindent\textbf{HW-based De-biasing:} In our application scenario, there are three design specific requirements: i)~minimize the de-biasing computational overhead on the token $\mathcal{T}$; ii)~avoid any additional data transmission overhead between the prover $\mathcal{P}$ and the token $\mathcal{T}$; and iii)~reduce additional de-biasing information stored on the token $\mathcal{T}$. We discuss how we achieve the first two goals in this section and explain how we achieve the third goal in Section~\ref{sec:Mblock}.

 
To achieve the first two goals, we offload the computational burden to the prover $\mathcal{P}$ by performing a one-time de-biasing during the enrollment phase. In the Pre-selection stage described in Section \ref{sec:preselection}, stable SRAM PUF bytes are identified. We further winnow HW balanced bytes from those stable bytes in our one-time de-biasing process. In general, we determine the address of bytes which are not only reliable but also HW balanced; HW close to 0.5. This approach leads to PUF responses---a linear combination of de-biased stable bytes---to be HW balanced. The HW-based process applied post pre-selection is shown in Figure~\ref{fig:preSelection}. 
\newline
\newline
\noindent\textit{\textbf{Remark:}} Our HW-based scheme does not deteriorate reliability of PUF responses, eschews leakage from potential bias and enables re-usability. As a trade-off, efficiency is decreased as the  $\frac{|\bf r_x|}{|\bf r_y|}$ is low as demonstrated by our experimental results in Appendix~\ref{app:selectioneff}. However, efficiency is not a concern in our implementation of SecuCode as there are always sufficient SRAM responses while we only need to use a small fraction of them. 
In~\cite{maes2015secure}, due to the de-biasing method---i.e. classic von Neumann de-biasing (CVN)---the token $\mathcal{T}$ generates different de-biasing data for each re-evaluation of the response ${\bf r_x}$. Thus, multiple observation of the de-biasing data given the regeneration of $\bf r_x$ leads to extra entropy loss. Therefore, the key generator based on such a de-biasing method is not reusable unless further optimization, e.g., pair-output VN de-biasing with erasures ($\epsilon$-2O-VN), is implemented to prevent extra entropy loss from the multiple production of the de-biasing data. In contrast, in our HW-based de-biasing method, the de-biasing data is generated only \textit{once} during the enrollment phase, we do not generate multiple de-biasing data and thus incurs no further entropy leakage from the de-biasing data. This implies that our HW-based de-biasing is reusable.

\subsubsection{On-chip meta-data storage structure}
\label{sec:Mblock}
It is inefficient to store the absolute addresses of winnowed bytes post pre-selection and de-biasing process, described in Figure~\ref{fig:preSelection}, on the CRFID device and to subsequently perform an exhaustive search to find them on demand. We propose the data storage structure named \textit{CRP-block map} to refer to those reliable and HW balanced SRAM PUF responses. 

The CRP-block map data structure is illustrated in Figure~\ref{fig:preSelection}(c). The CRP-block data structure divides the SRAM memory into multiple blocks. Each block has an integer index such as Block 0, Block 1, and Block 2. The size of each block may vary, however, each block is designed to contain an equal number of reliable and HW balanced bytes required for the physically obfuscated key derivation mechanism. To map a CRP-block into a physical memory address, each block has an absolute starting address and several offsets pointing to the winnowed bytes. The starting address of each block and the offsets are stored in a look up table (LUT) in a token's NVM indexable via block number. To produce a PUF key, the \textsf{TRNG} generates a seed challenge that is a block number. Consequently, a look up of a block number will resolve the targeted block's starting memory address and the offsets point to the response bytes. Hence a random challenge ${\bf c}_i$ can be used to select a CRP-block and, subsequently, the power-up states of the bytes in the selected CRP-block are readout and concatenated as the response ${\bf r}_i$ that is both highly reliable and unbiased.

\subsection{Message Authentication Code}
SecuCode requires one cryptographic primitives: a keyed hash function to build a message authentication code (MAC) to realize the \textsf{MAC()} function. The two dominant factors determining their selection are:
\begin{enumerate}
\item The CPU clock cycles required for execution. For example, we can see from Fig.~\ref{fig:availableCycle} that at a 50~cm distance, we may not expect more than 500,000 to 600,000 clock cycles before harvested power is exhausted. 
\item Memory (RAM) budget. Since available on-chip memory is shared by the RFID communication stack, sensor data, user code, and SecuCode, the state space available for a cipher execution is limited.
\end{enumerate}

\textcolor{T}{Therefore, power and memory efficient primitives are highly desirable. The construction of a \textit{secure}, computation and  power efficient message authentication code~\textsf{MAC()} on the token is extremely challenging as the entropy compression task is resource intensive and attacks, such as birthday attacks, demand that we use a MAC function with a large enough output size.} 

\textcolor{T}{Given the lack of MAC benchmark data for MSP430 microcontroller series, we selected and implemented a set of existing secure keyed hash functions expected to yield a computationally efficient software implementation. We implemented and evaluated: BLAKE2s-256, BLAKE2s-128, and HWAES-GMAC and HWAES-CMAC for comparison; here \textit{HWAES-} functions benefited from the AES hardware accelerator module on the MSP430 MCU. The results of our study are detailed in Appendix~\ref{Appendix:mac}. Based on these results, we selected a 128 bit MAC using \textit{HWAES-CMAC}, Cipher-based Message Authentication Code~\footnote{We used the implementation detailed in NIST Special Publication 800-38B, \textit{Recommendation for Block Cipher Modes of Operation: the CMAC Mode for Authentication}} built using AES since it yields the lowest clock cycles per byte.}

\subsection{Intermittent Execution Model}
\begin{figure}
\centering
\includegraphics[width=\linewidth]{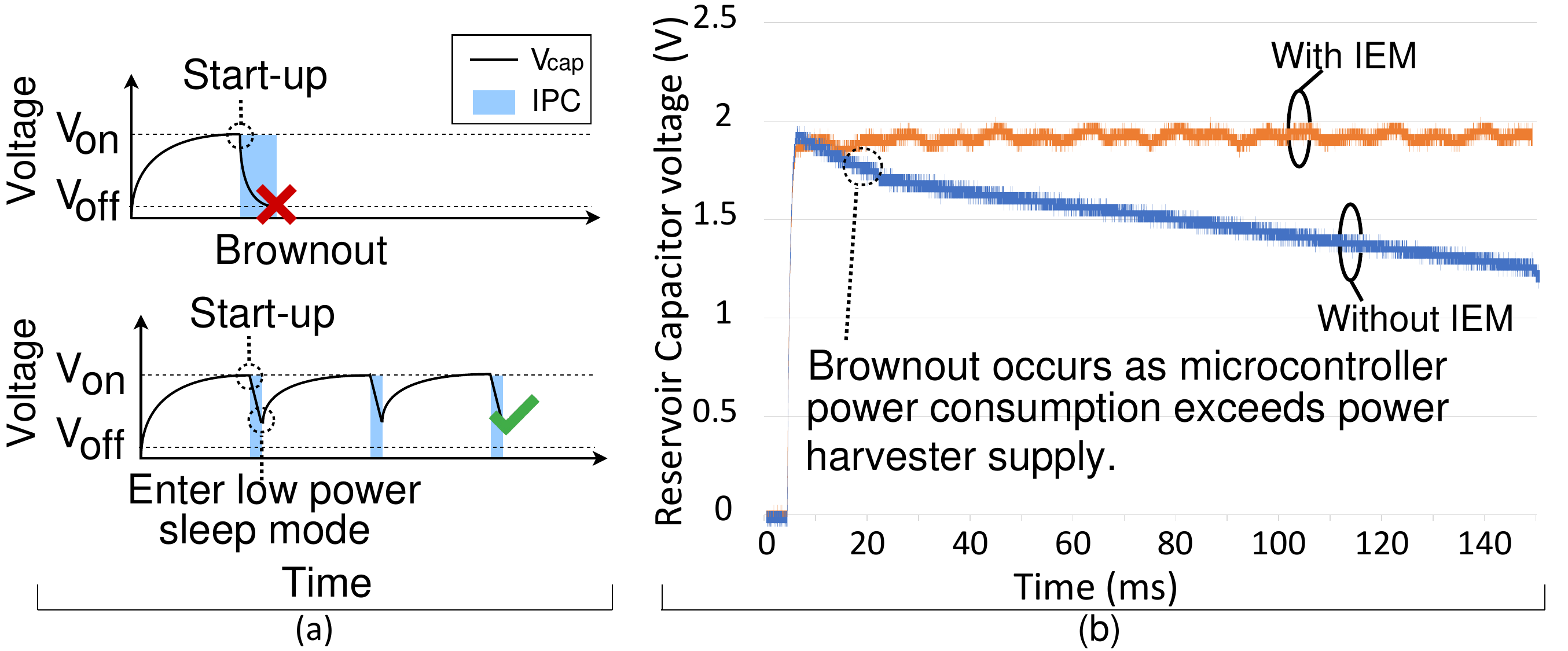}
\textcolor{T}{\caption{(a) Illustration of a brownout event and the proposed intermittent execution model (IEM) to prevent brownouts during computationally intensive operations (notations are given in Fig. \ref{fig:suddenPowerLoss}). (b) Comparison of a PUF key derivation on a CRFID transponder at 50~cm from an RFID reader antenna with and without our proposed IEM.}}
\label{fig:IEM-measure}
\end{figure}

The nature of intermittently powered devices that rely on harvested power is typically described by a power harvesting and charging phase where energy is generally stored in a reservoir capacitor and then released for powering computations. We refer to this cycle as the IPC---Intermittent Power Cycle. \textcolor{T}{A brownout event can occur when the available energy in an IPC subceeds the energy needs of the computations, and the power harvester is unable to replenish energy as rapidly as it is consumed. A brownout results in state loss and termination of the execution thread as highlighted in Fig~\ref{fig:suddenPowerLoss}. Hence, as illustrated in Fig.~\ref{fig:IEM-measure}, we cannot always expect to continue a computation task to completion.} 

\textcolor{T}{
Studies such as Alpaca~\cite{Maeng:2017:Alpaca}, Mementos~\cite{ransford2011mementos}, CCCP~\cite{salajegheh2009cccp} and Clank~\cite{hicks2017clank} consider the problem of continuing the thread of execution over periods of power loss.
In general, these studies employ checkpoint based methods with various degrees of programmer support. The basic concept is to save state---checkpoint---and to restore state to a previously valid checkpoint to allow the resumption from a previous state of execution after a power loss event. With the exception of CCCP, state is saved in a device's non-volatile memory; CCCP proposes the saving of state in an untrusted server. However, in SecuCode, process state such as the PUF key is intentionally volatile to eliminate the need to protect key storage. If power-loss or brownout does occur, all unfinished process state based on old key material should be discarded. Therefore, approaches to save and restore state from non-volatile sources are not desirable. Further, writing to NVM is energy intensive and we would like to avoid the additional overhead of checkpointing.} \textcolor{T}{In contrast, Dewdrop~\cite{buettner2011dewdrop} considers execution under frequent power loss by attempting to \textit{prevent} a brownout event by solving a task scheduling problem; execute tasks only when they are likely to succeed by monitoring the available harvested power. Dewdrop provides an elegant dynamic scheduling method, however, requires the overhead of sampling the harvester voltage within the application code and task scheduling.} 

\textcolor{T}{In order to deal to with frequent intermittent power loss, we consider the following intermittent execution model (IEM) for computation intensive building blocks of the protocol: \textsf{MAC()} and \textsf{FE.Gen()} functions. Our IEM is built on the basic concept of a \textit{task} in Alpaca~\cite{Maeng:2017:Alpaca}---a code block proportioned to execute to completion under a minimum number of available clock cycles or energy--- and the concept in Dewdrop~\cite{buettner2011dewdrop} attempting to \textit{prevent} brownouts. We first identify the computationally intensive functions in the firmware, and construct an execution plan based on factoring the function to subtasks. These subtasks are then interleaved with low power sleep states in the code.  We selected the lowest power consuming microcontroller sleep state from the target microcontroller such that the memory state is maintained during sleep. The duration of the imposed sleep state---termed the \textit{intermittent operating setting}---is realized using an on-chip timer interrupt to wake up the transponder at the conclusion of a given intermittent operating setting; upon wake-up the transponder continues the execution of the following subtask.} 

\textcolor{T}{Immediate benefit of the IEM is the possible prevention of an intentional brownout event by allowing the replenishment of the reservoir capacitor energy by the power harvesting and charging circuity as illustrated by the graph at the bottom of Fig.~\ref{fig:IEM-measure}(a).  Fig. \ref{fig:IEM-measure}(b), shows an experimental validation of of our IEM; here, we plot two captures of reservoir capacitor voltage with and without the IEM. The voltage capture traces show that the IEM restores power during low-power sleep states interleaved between subtasks---shown by the recovery of the voltage developed across the reservoir capacitor. In contrast, as shown by a falling reservoir capacitor voltage, the CRFID device without IEM  fails due to brownout. Here, the CRFID device is unable to harvest adequate power to replenished the reservoir capacitor fast enough to provide the minimum operating voltage (1.8~V) necessary for the MCU. }


\subsection{End-to-end Implementation}\label{sec:bootloader}
\begin{figure*}[!ht]
    \centering
    \includegraphics[width=0.8\linewidth]{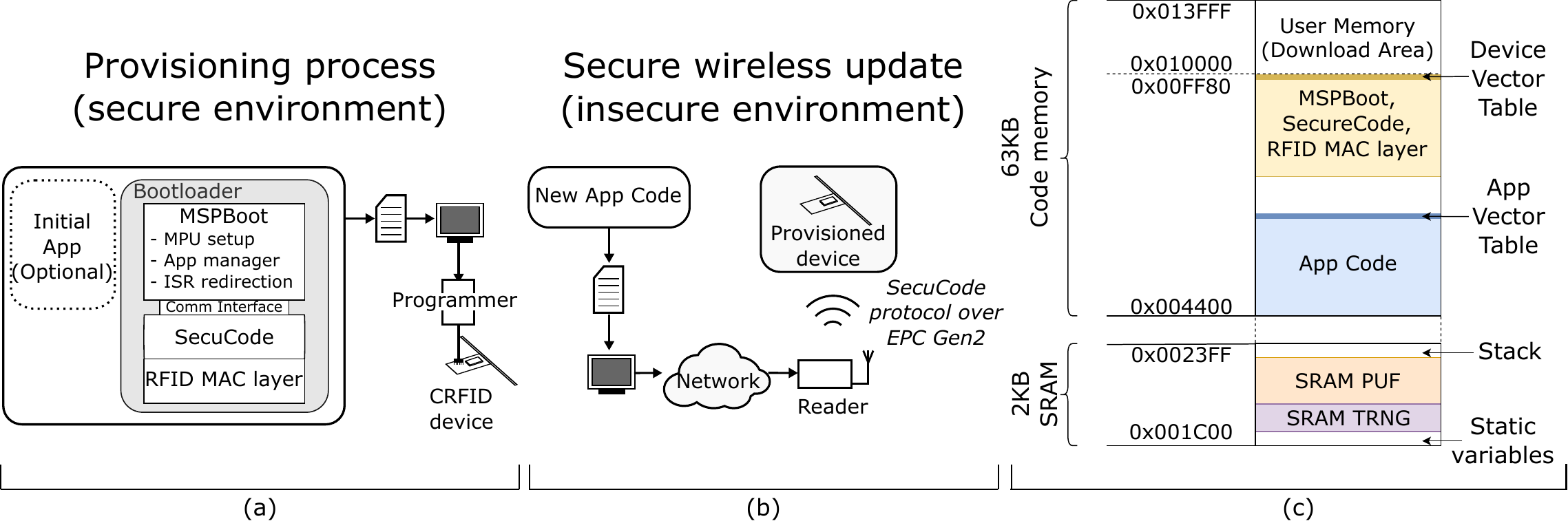}
    \caption{(a) Overview of bootloader provisioning; (b) Secure wireless firmware update; and (c) Memory arrangement for the 2~KB SRAM and 64~KB FRAM.}
    \label{fig:OTAfirmware}
\end{figure*}
The complete SecuCode based firmware update process is illustrated in Fig. ~\ref{fig:OTAfirmware}. We describe below the memory arrangement, the development of the bootloader together with the complete tool chain to realize a standards compliant secure firmware update process. 

\vspace{0.2cm}
\noindent\textbf{Memory Arrangement:} As shown in Fig.~\ref{fig:OTAfirmware} (c), the 2~KB SRAM embedded in the MSP430FR5969 MCU is divided into two sections. The lower address space is used for the SRAM TRNG and occupies 64 words of SRAM. The address space above the SRAM TRNG forms the SRAM PUF. The SRAM PUF and SRAM TRNG do not span the full space of the SRAM memory. This is necessary to allocate space for initialization routines and PUF state variables. One hundred and sixty bytes of higher addresses are allocated as stack space, and 480 bytes from the lower address space is designated for static variables such as ${\bf c_i}$, $nonce$ and ${\bf h_i}$. 
The SRAM PUF and SRAM TRNG are only active during the \textbf{Physically Obfuscated Key Derivation} stage in Fig.~\ref{fig:Flowchart}; once the response ${\bf r}_i$ is readout and ${\bf sk_i}$ and ${\bf h_i}$ are generated, the SRAM memory is released for regular operations.

The CRFID device we employed has an embedded 63~KB Ferroelectric Random Access Memory (FRAM) as NVM. FRAM is partitioned into 63~KB code memory and 1~KB for Device Descriptor Info~\cite{instrumentsmsp430fr59xx}. In this work, we only employ the 63~KB code memory space. The FRAM memory layout of the specific CRFID device we used is shown in Fig. \ref{fig:OTAfirmware}(c). FRAM is divided into three sections: i) the bootloader; ii) Application code, and iii) User Memory (Download Area). The User Memory segment can be manipulated using \texttt{BlockWrite} commands, however only the bootloader can write to the Application Code memory space.

\vspace{0.2cm}
\noindent\textbf{bootloader:}~The SecuCode implementation for the CRFID device requires a bootloader to be provisioned onto the device. This is because the CRFID device has no supervisory control of an operating system and is designed in the manner of a low-end and low-cost device.

Our bootloader is based on TI's bootloader framework, MSPBoot~\cite{ryanbrownkatiepier2016}. The \texttt{Comm} interface is designed to operate on trusted data and therefore should not directly communicate with the RFID Media Access Control layer. Instead, the received firmware is buffered in a Download Area in memory. After the SecuCode protocol verifies the authenticity and the integrity of the firmware, it is passed to the MSPBoot via the \texttt{Comm} interface. In our implementation, we adopted the WISP5 firmware\footnote{\url{https://github.com/wisp}} for the RFID Media Access Control layer.

\textcolor{T}{In order to realize an immutable bootloader, we considered the full memory protect mode (FMPM) and partial memory protect mode (PMPM) offered by the microcontroller together with setting an e-Fuse~\cite{pier2015msp} to disable the potential for wired re-programming using the on-chip JTAG interface after deployment in the field.} In FMPM, the MPU (memory protection unit) is configured to prevent writing to the bootloader. Most importantly, the MPU is locked from being accessed. These actions are performed during the initialization process, before execution of user code, and therefore it is infeasible for the bootloader to be modified in FMPM mode. In PMPM, writing to the bootloader is still prevented, however, the MPU is not locked. Such an MPU configuration provides basic protection for the bootloader against programming errors, while still allowing the bootloader to be remotely updated. In this context, the firmware temporarily disables memory protection, overrides parts of the original code, then re-enables memory protection. We employ FMPM mode to realize an immutable bootloader.

\vspace{0.2cm}
\noindent\textbf{Secure Firmware Update Process:}~The provisioning and the secure wireless update process is illustrated in Fig.~\ref{fig:OTAfirmware}. 
A CRFID device is first provisioned whereby the immutable program called the \textbf{bootloader} is installed on the device in a secure environment. 
We assume the wired interface is disabled after the installation of the bootloader 
and therefore wireless code dissemination is the only practical mechanism by which to alter the firmware. 
The provisioned CRFID device is deployed in the field and can be subsequently updated with new firmware following the process below.
\begin{itemize}
\item Compile the new firmware using a standard MSP430 compiler and pass the output, together with a linker map that specifies the memory allocation scheme shown in Figure \ref{fig:OTAfirmware}, to an MSP430 linker to generate a binary file in ELF format\footnote{Alternatively, Texas Instrument's Code Composer Studio integrated development environment which bundles all the necessary tools can also be used to simplify the task}.

\item Use our SecuCode App, available from~\cite{secucodegithub}, to load and parse the resulting ELF file to wrap the \texttt{LOAD} segments into MSPBoot commands specified in the MSPBoot framework \cite{ryanbrownkatiepier2016}. The resulting binary is broken up into 128-bit blocks for transmission to a CRFID transponder.

\item The SecuCode App subsequently uses LLRP commands to construct \texttt{AccessSpecs} and \texttt{ROSpecs}---refer to Section~\ref{sec:c1g2} for LLRP. These encodes \epc protocol commands employed by SecuCode such as \texttt{Authenticate}, \texttt{BlockWrite} and \texttt{SecuComm} commands\footnote{Given that Impinj R420 RFID readers do not yet support the recent \epc protocol changes, we implement the unsupported commands using \texttt{BlockWrite} commands. The implementation of the employed commands are detailed in Appendix.\ref{sec:commandCoding}} to setup a networked RFID reader to execute the SecuCode protocol to wirelessly and securely update the firmware of the target device.

\end{itemize}


\section{Experimental Results and Analysis}\label{sec:experiments}
In this section, we describe the extensive set of experiments conducted to evaluate SecuCode. In our work, we employed 20 MSP430FR5969 MCU chips and a WIPS5.1LRG CRFID device built with the same microcontroller. The MSP430FR5969 microcontrollers consists of 2 KB (16,384 bits) of internal SRAM memory and 64~KB of internal FRAM memory. In our implementation we employed an 8-bit challenge ${\bf c_i}$ and a 128-bit $nonce$. We summarize our experiments below: 

\begin{enumerate}
\item We evaluate and validate the response pre-selection and HW-based de-biasing methods, and subsequently, determine the BER of the CRP block maps to determine the parameters $n$, $k$, $t$ for the BCH encoder used to realize the \textsf{FE.Gen} function on the CRFID transponder (Section~\ref{sec:ExReliability}).

\item \textcolor{T}{Given that: i) security protocol implementation costs are rarely examined in the literature; and ii) the resource constrained nature of the CRFID devices demand security, performance as well as practicability; we evaluate the memory requirements (code size in FRAM and state space in SRAM) and computing clock cycles for the security functional blocks (those in addition to MSPBoot and RFID MAC Layer) necessary for SecuCode (Section~\ref{sec:ExReliability} and \ref{sec:footprint}).}


\item \textcolor{T}{We evaluate the impact of the most energy and computationally demanding SecuCode functional blocks---key derivation and MAC function---on performance and evaluate the effectiveness of our intermittent execution model on these energy intensive building blocks. (Section~\ref{sec:exp_coldstart_overhead} and Section~\ref{sec:exp_hash}).} 


\item We present a case study to demonstrate the end-to-end implementation of SecuCode in an application scenario---the source code of our case examples are released on our project website to facilitate further research, experimentation and adoption of our SecuCode scheme (Section~\ref{sec:casestudy}). 
\item We evaluate the security of SecuCode under the adversary model detailed in Section~\ref{sec:adversarymodel} (Section~\ref{sec:securityAnalysis}).
\end{enumerate}
\begin{figure}[!b]
  \centering
  \begin{minipage}{0.54\linewidth}
   \vspace{5pt}

\includegraphics[width=\linewidth]{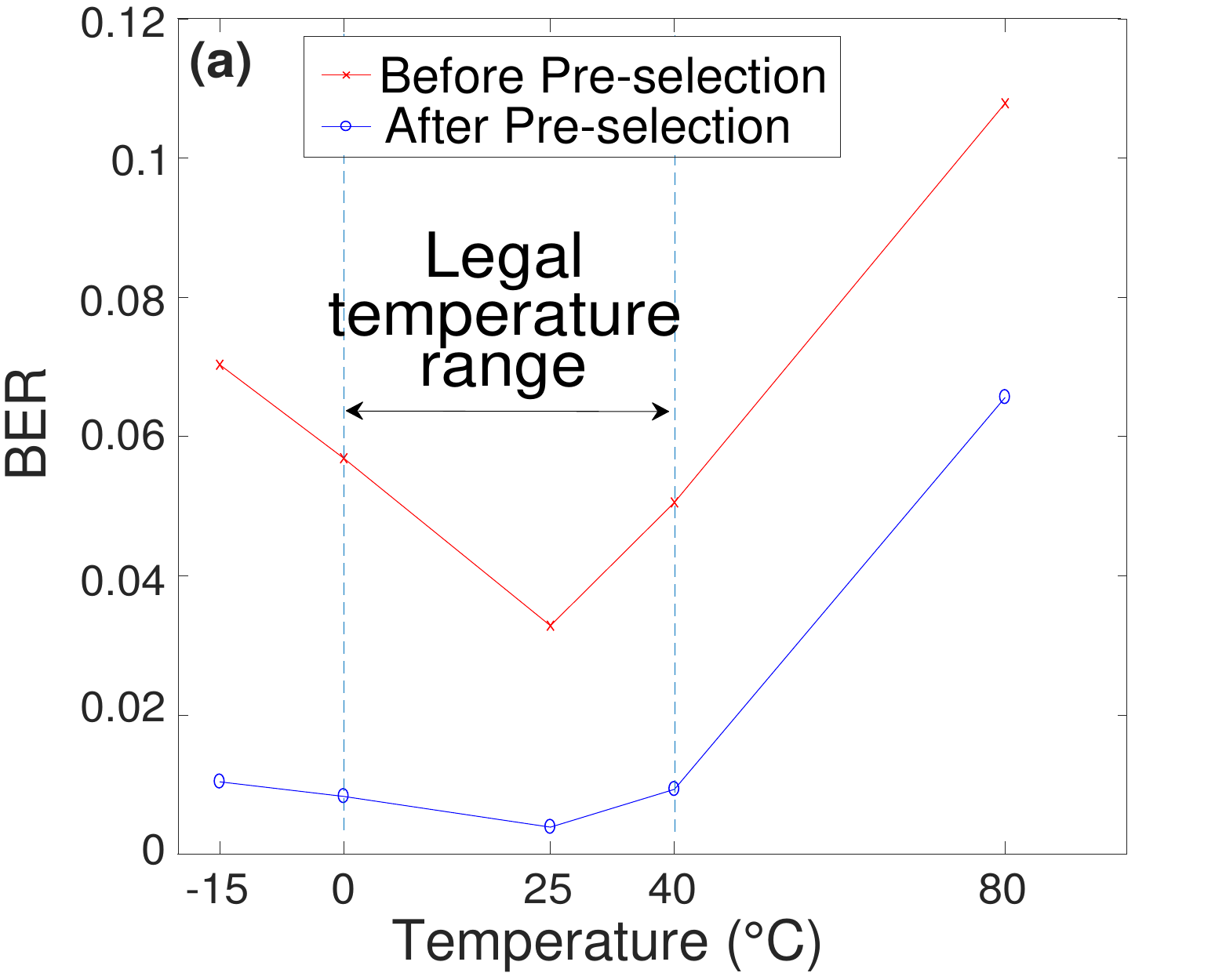}
  \end{minipage}
  \begin{minipage}{0.44\linewidth} 
  \includegraphics[width=\linewidth]{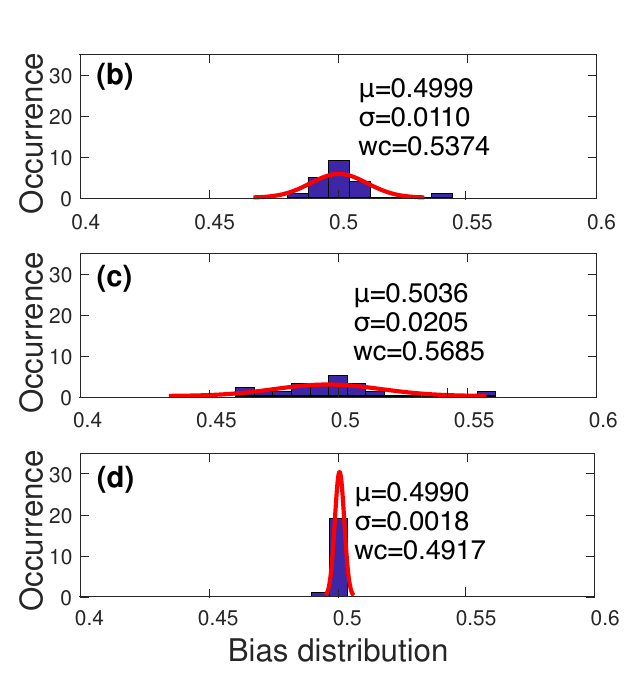}
   \end{minipage}
\caption{(a) BER across differing operating temperatures. Bias distribution of 20 tested chips:~(b)~raw PUF responses; (c)~after pre-selection; and (d)~after HW-based de-biasing.}
\label{fig:BER_temp}
\label{fig:Mblocks-Debias}
\end{figure}

\subsection{Physically Obfuscated Key Derivation}\label{sec:ExReliability}
In this section, we validate the lightweight physically obfuscated key derivation method using experimental data.
\newline
\newline
\noindent\textbf{SRAM PUF Reliability:} Considering that SRAM PUFs are insensitive to voltage variations~\cite{selimis2011evaluation} and the voltage can be eventually controlled well in practice, we focus on its performance under differing temperature corners (-15°C, 0°C, 25°C, 40°C and 80°C). Under each temperature corner, we repeat the response readout 100 times. Each SRAM PUF can generate 16,384 response bits, we tested 20 such PUFs. The BER values for various corner temperatures is shown in Fig.~\ref{fig:BER_temp}(a). The applied pre-selection reduces the BER significantly to no more than 0.94\% over the defined legal temperature range of 0°C to 40°C---see Section~\ref{sec:preselection}.
\newline
\newline
\noindent\textbf{SRAM PUF Bias:} We follow the method in~\cite{delvaux2016efficient} to evaluate PUF bias. The bias distribution $b$ is expressed as:
\begin{equation}
\label{eqn:bias}
	{\rm b} = \mathbb{P}({\bf r_i}=1)~~\textnormal{for}~i\in \left[1,n\right]
\end{equation}
where $\bf r$ is a response vector of length $n$.
After response pre-selection, we implemented our HW-based de-biasing method. As shown in Fig.~\ref{fig:Mblocks-Debias}(d), experimental results from the 20 SRAM PUFs show that the mean of the bias is 0.499; very close to the ideal value of $b=0.5$.
\newline
\newline
\noindent\textbf{Reverse Fuzzy Extractor using BCH($n,k,t$):}
Although our BER is less than 1\%, as experimentally evaluated and shown in Fig.~\ref{fig:BER_temp}, we selected to evaluate: i) 8 blocks of BCH(31,6,3) code capable of correcting up to $\frac{3}{31}$ or 10\% of bits
 with $P_{\rm Fail}=0.0016$ for 8 parallel blocks; and ii) 5 blocks of BCH(63,24,7) code capable of correcting up to $\frac{7}{63}$ or 11\% of bits 
with $P_{\rm Fail}=7.4506\times10^{-7}$ for 5 parallel blocks according to~ Eq.~\eqref{eq:p-fail-for-parallel}. 
Executing eight BCH(31,16,3) blocks to derive a 128-bit key requires 146,535 clock cycles while computing 5 blocks of BCH(63,24,7) to obtain a 120-bit key consumes 288,980 clock cycles. 

Although employing a BCH(63,24,7) code allows us to achieve an industry standard key failure rate of less than $10^{-6}$, we will see in Section~\ref{sec:exp-end-to-end} that protocol failure due to external conditions such as intermittent powering is more likely than protocol failure due to a failed key recovery.
Therefore, we employed parallel blocks of BCH(31,16,3) code as a compromise between error correction capability, computational complexity and performance in practice. 
In general, for each block, the computational complexity of finding the 31-bit response ${\bf r_i}$ through the corresponding 15-bit public $\bf h$ is $2^{16}$---assuming that the response bits are uniformly distributed and have no correlations with each other. In order to achieve a 128-bit security level with an attack complexity of $2^{128}$, we need eight such blocks as discussed in Section~\ref{sec:preselection}. Thus, an SRAM PUF response ${\bf r_i}$ with $31\times 8=248$ bits need to be readout from the SRAM PUF for each protocol session.


\textcolor{T}{\subsection{SecuCode Implementation Footprint}\label{sec:footprint}
In Table~\ref{tab:MemNClk}, we summarize the cost, in terms of CPU clock cycles, FRAM memory usage for code size and SRAM memory usage for state space size, for implementing each security related building block of SecuCode on a CRFID transponder. These blocks represent the necessary overhead imposed on the CRFID transponder to build our secure update method to prevent malicious code injections attacks.
While the computational cost of the reverse fuzzy extractor is significant, it is fixed for each iteration of an update session. In contrast, the computational cost of the \textsf{MAC()} function can increase with larger firmware code blocks (see our evaluations in Appendix~\ref{Appendix:mac}).}
\begin{table}[!t]
	\centering
	\caption{Memory \& execution load of functional blocks}
	\label{tab:MemNClk}
    \resizebox{\columnwidth}{!}{%
        \begin{tabular}{ccc|c}
        \toprule[1.5pt]
        \multirow{2}{*}{Protocol Steps} & \multicolumn{2}{c|}{Memory Footprint (Byte)}  & \multirow{2}{*}{Clock Cycles} \\ \cmidrule{2-3}
        & Data(SRAM) & Code(FRAM) &  \\\hline
        $nonce ←$\textsf{Rand}() & 6 & 68 & 375\\
        ${\bf r_i}$ $←${\sf PUF}(${\bf c_i}$) & 6 & 32 & 615\\
        OTF ← \textsf{Temp()} & 10 & 204 & 734\\
        ${\bf h_i}$ $←$ \textsf{FE.GEN}(${\bf r_i}$)\footnotemark[1] & 53 & 621 & 109,234\\
    $s^{\prime}$ ← {\sf MAC}(firmware$\|nonce$,${\bf sk_i}$)\footnotemark[2] & 58 & 3,198 & 22,197\footnotemark[3] \\
        \hline
        SecuCode total & 133 & 4,123 & 133,155 \\\bottomrule[1.5pt]\\ 
        
        \end{tabular}
    }
    \raggedright 
    \footnotemark[1]{\textsf{FE.GEN} based on BCH(31,16,3) code.\\}
    \footnotemark[2]{using a 128-bit HWAES-CMAC message authentication code.\\}
    \footnotemark[3]{For a moderate firmware size of 240 bytes.}
\end{table}

\subsection{SecuCode Overhead and IEM Settings}\label{sec:exp-end-to-end}
\textcolor{T}{Table~\ref{tab:MemNClk} shows us that the most computationally intensive security building block are the \textsf{FE.Gen} (key derivation building block) and \textsf{MAC} (needed for firmware integrity checks and prove authentication) implementations. Hence these building blocks are likely to be most impacted by as well as be the cause for brownout events. Therefore, both \textsf{FE.Gen} and \textsf{MAC} implementations operate under our IEM.}

\textcolor{T}{A cold start-up is when a CRFID tag is activated from a completely powered down sate---i.e. where the reservoir capacitor of the CRFID device has been discharged in the absence of a wireless powering source. Cold start time is defined as the time taken by a CRFID device to respond to an RFID reader inventory command from a cold start-up. As illustrated in the control flow diagram in Fig. \ref{fig:Flowchart}, physically obfuscated key derivation steps occur after a cold start-up. Therefore, the \textsf{FE.Gen} or key derivation method will place additional energy and computation burdens on a cold start-up initialization process and potentially lead to unsuccessful device start-ups at the initiation of a firmware update. Hence, we evaluate the \textit{cold start-up success and time overhead} of a key derivation operation.}

\textcolor{T}{We also evaluate the success rate of our \textsf{FE.Gen} and \textsf{MAC} functions under wireless powering conditions---different distances from a wireless powering source, i.e. RFID reader antenna---and intermittent operating settings to understand the overhead imposed by our IEM as well as the effectiveness of IEM to mitigate brownouts. We detail our experimental method in Appendix~\ref{Appendix:expSetup}.}

\subsubsection{Physically obfuscated key derivation}\label{sec:exp_coldstart_overhead}

\begin{figure}[t]
\centering
\includegraphics[width=\linewidth]{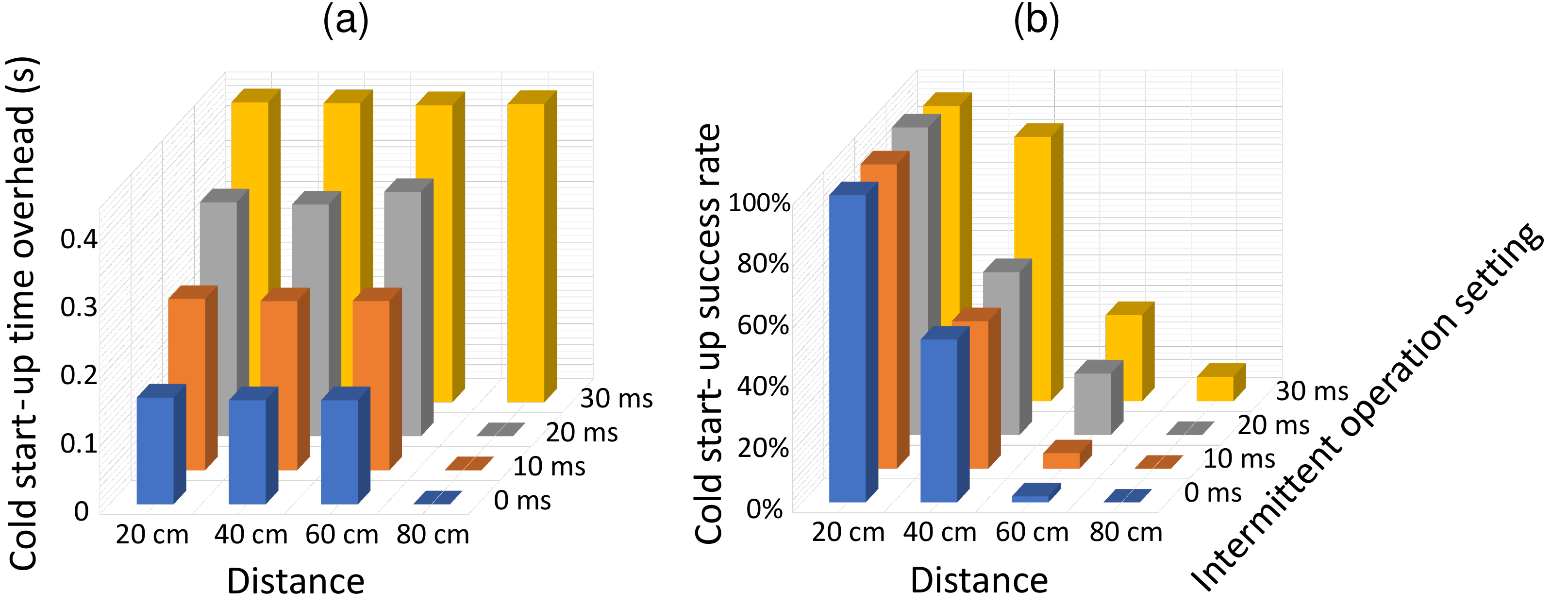}
\caption{Physically obfuscated key derivation: (a) Cold start-up time overhead and (b) success rate. }
\label{fig:TimePowerOverhead}
\end{figure}

In this experimental setting, we place the CRFID transponder (WISP5.1-LGR) provisioned with our bootloader at 20~cm, 40~cm, 60~cm and 80~cm apart from a 9~dBi circularly polarized reader antenna oriented towards a high ceiling to minimize interference from multipath signals on our observations. We used each of the eight BCH code computations as subtasks for our intermittent execution model (IEM).
    
The results of cold start-up times and success rates are plotted in Fig.~\ref{fig:TimePowerOverhead}(a) and (b), respectively; here we plot mean cold start-up times and success rates over 100 repeated measurements collected for each intermittent operating setting and distance pair. At 20~cm interrogation range, the CRFID transponder completes key derivation from a cold start-up with a high success rate; approximately 100\%. When the distance is 40~cm, the success rate witnessed a dramatic drop down to less than 50\%, unless we increase the intermittent operation setting to 30~ms. When the distance is beyond 60~cm, the success rate cannot be guaranteed even using a large intermittent operation setting. Achieving higher success rates beyond 60~cm will require further division of subtasks capable of utilizing the available clock cycles before brownout. Most importantly, we observe that the IEM can increase the success rate of completing the key derivation process from a cold start-up but the trade-off is an increase in the cold start-up time overhead as illustrated in see Fig. \ref{fig:TimePowerOverhead}(a).

\textcolor{T}{\subsubsection{MAC Computation}\label{sec:exp_hash}
We define \textit{MAC function latency} as the time interval between calling a MAC function to perform a calculation over a block of pre-loaded data on the CRFID transponder and the generation of a valid result. We employ a single data block size. Following the definition in~\cite{eisenbarth2010evaluation}, the data size is large enough such that the latency is mainly dominated by multiple compression rounds. In this experiment, a random string of 1,280 bytes is used. Notably 1,280 bytes is twice the size of the largest firmware update used in our experiments. Since the hardware AES accelerator is very efficient in terms of clock cycles, instead of considering each AES compression round as a subtask, we employ 32 rounds of AES computations as a subtask in our IEM implementation.}

\begin{figure}
\centering
\includegraphics[width=\linewidth]{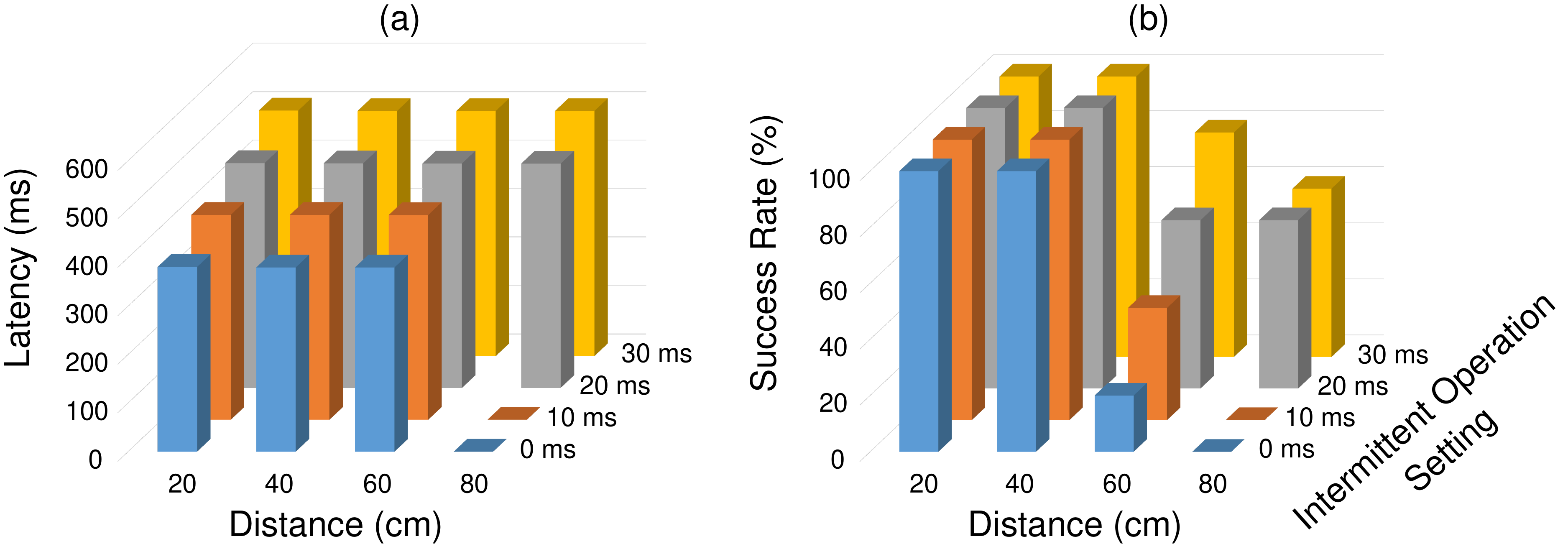}
\caption{(a) MAC function success rate and (b) success rate at different distances from an RFID reader antenna under different intermittent operation settings. With data size 1280 bytes.}
\label{fig:hashSuccessRate}
\end{figure}

\textcolor{T}{We report the mean latency and success rate obtained over 10 repeated MAC computations collected for each intermittent operating setting and distance pair in Fig.~\ref{fig:hashSuccessRate}. 
As expected, the observed latency increases linearly along with the intermittent operating setting. Overall, increasing the intermittent operation setting improves the mean success rate of the \textsf{MAC} computation. For example, in Fig.~\ref{fig:hashSuccessRate}(b), we can see that at a 60~cm distance, the success rate of the MAC  computation improves from 20\% to 80\% with an intermittent operating setting of 30~ms. Further, we can achieve success of no less than 60\% at 80 cm distance when the intermittent operating setting is greater than 20~ms.}

\begin{figure}[!ht]
\centering
\includegraphics[width=.65\linewidth]{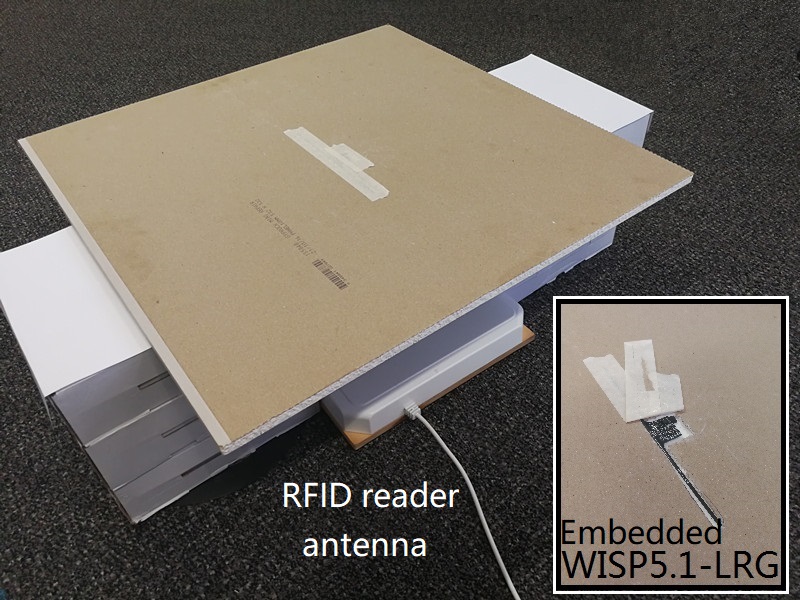}
\caption{SecuCode case study set-up. The CRFID transponder is embedded in a plaster tile and placed 20~cm above the antenna to emulate the target application scenario.}
\label{fig:plaster}
\end{figure}

\begin{table}[!ht]
	\centering
    \caption{\textcolor{T}{SecuCode case study performance measures.}}
    \label{tab:endToEnd}
    \begin{tabular}{C{1.0in}|C{.4in}C{.8in}C{.5in}}
    \toprule[1.5pt]
    Application code & Size (Bytes) & Latency (mean, std)\footnotemark[1](s) & success rate (\%)\\\hline
    Accelerometer service code&399 & 3.1, 1.0 & 81 \\\hline
    Thermometer service code&273 & 2.5, 0.6 & 92  \\\hline
    \vspace{3pt}
    GLN identifier code&  223 &  2.1, 0.5 &  90\\
    \bottomrule[1.5pt]
    \multicolumn{3}{c}{}\\
    \end{tabular}
    \raggedright
    \footnotemark[1]{\textcolor{T}{Latency was measured as the time interval between the SecuCode App transmitting the firmware to an RFID reader using LLRP commands and the time SecuCode App confirms the acknowledgement of a successfully updated firmware.}}
\end{table}

\subsubsection{SecuCode Case Study}\label{sec:casestudy} 

As a demonstration of SecuCode in an application scenario, we evaluate SecuCode in the following setting: \textit{There are several CRFID transponders embedded in plasterboards mounted as ceiling tiles in a chemical warehouse. We require three different types of services from these transponders. Some are to be programmed with \textit{Accelerometer service code} to monitor potential structural failures, some are to be programmed with \textit{Thermometer service code} to detect potentially dangerous thermal storage conditions while others are to be programmed with basic firmware to respond with a fixed  \textit{Global Location Number (GLN)} as anchor points to identify storage locations. The embedded CRFID transponders need to be reprogrammed wirelessly to support monitoring and location service needs of the warehouse; further, over time, changes to the warehouse layout and monitoring needs such as hazardous temperature levels can require alterations to existing firmware.}

As shown in Fig. \ref{fig:plaster}, to emulate such a scenario, we embedded a WISP5.1-LGR into a plaster tile after provisioning the device with our bootloader. The plaster tile with the embedded WISP is then placed 20~cm above an RFID reader antenna. We used an IEM setting of 30~ms as it provided a good compromise between latency (time to complete a firmware update) and a high success rate over a range of wireless transfer distances.  The results from 100 repeated wireless firmware updates based on the three firmwares specific to three distinct applications in our scenario description is summarized in Table \ref{tab:endToEnd}. We successfully reprogrammed all three application firmwares within a few seconds. Most notably we observed an increased number of failures due to brownout for the larger firmware owing predominately to the increased energy required for the \textsf{FE.Gen} computation. Further, as expected, all of the failures were due to brownout as opposed key recovery---recall that the key failure rate for the BCH code used is less than 0.2\% as shown in Section~\ref{sec:ExReliability}. Demonstration video of the firmware update process and the source code of SecuCode App are available from\cite{secucodegithub}.

\subsection{Summary}
Our results demonstrate: i) the practicability and robustness of the developed SecuCode protocol; ii) the relationship between operating conditions, intermittent operation settings and protocol performances; and iii) the effectiveness of the proposed IEM on realizing successful secure firmware updates under intermittent powering from harvested energy.  


\subsection{Security Analysis}
\label{sec:securityAnalysis}
We analyze the security of SecuCode under the adversary model detailed in Section~\ref{sec:adversarymodel} in the following. 

\vspace{0.2cm}
\noindent{\bf Man-in-the-middle attack:} Here, security is related to the complexity of the adversary $ \mathcal A $ fooling the token $ \mathcal T $ in order to inject malicious code. Such an adversary is faced with the task of determining the secret key ${\bf sk_i}$ and therefore, we evaluate the security of the reverse fuzzy extractor which leaks helper data during a firmware update session. 

We employ a reverse fuzzy extractor where the helper data computation is placed on the resource-constrained token $ \mathcal T $.  In this context, different helper data corresponding to the same response $ {\bf r_i} $ will be generated during different firmware update sessions; thus, multiple helper data are exposed. It has been proved that the entropy leakage from the helper data is independent of the number of enrollments for SRAM PUF~\cite{kusters2017security}. In other words, one can evaluate the entropy leakage from the helper data of the reverse fuzzy extractor for a single helper data generation case; or the entropy leakage from multiple helper data observations is same as the entropy leakage from a single helper data observation, as with a traditional fuzzy extractor. \textcolor{T}{Further, we assume that the debiasing mask does not leak information. This is a reasonable assumption as the mask is limited to indicating the reliable CRP-blocks only and not the response itself. In addition, it has been demonstrated through extensive experiments that modern SRAM PUFs possess good uniqueness properties~\cite{maes2016physically}. In other words, the highly reliable bits are not correlated across different instances of SRAM PUFs. Therefore, knowledge of an SRAM PUF instance's mask can safely be assumed to not leak responses information of other SRAM PUFs.}

Then, given the  BCH($n$,$k$,$t$) code used to implement the reverse fuzzy extractor helper data computation, no more than ($n-k$)-bit entropy is leaked. To be precise, the residual min entropy~\cite{delvaux2016efficient} of the n-bit response is given by $ H_{\rm min}=-n\cdot\log_2(max(b,1-b))-(n-k)$ when the helper data is public, where the PUF response bias---the probability of a single PUF response being `0'--- is $ b $. Ideally, the bias $ b $ is 50\%. 
However, as we have seen, PUF responses can display a slight bias. Therefore to guarantee the ($n-k$)-bit entropy leakage 
we employed the de-biasing method proposed in Section.\ref{sec:HWde-biasing}.


As we use 8 blocks of BCH(31,16,3) code, we achieve a key ${\bf sk_i}$ with 128-bit entropy---strictly, 127.6 bits entropy considering the experimentally evaluated bias of 0.4990 post our de-biasing method. Therefore, without knowledge of $ {\bf sk_i} $, the probability of fooling the token $ \mathcal T $ to update a malicious firmware by $ \mathcal A $ is no more than the brute-force attack probability of $ 2^{-127.6} $.

\vspace{0.2cm}
\noindent{\bf Helper data manipulation attack.} We are aware of helper data manipulation attacks based on exposed helper data reported in~\cite{delvaux2015helper,beckerrobust}. However, both works acknowledge that such an attack is error correction code dependent and cannot be mounted on a linear code such as BCH employed in SecuCode~\cite{delvaux2015helper,beckerrobust}. Further, mounting such an attack on SecuCode is difficult; in the first place, we notice that such a helper data manipulation attack is very easy to be detected by the prover $ \mathcal P $. The reasons are as follows:
\begin{enumerate}
\item The frequency of normal firmware updates is very low in comparison with other services that are built with reverse fuzzy extractors, such as authentication or attestation services. However, helper data manipulation attacks require a large number of queries. Recall, that helper data is generated by the token $ \mathcal T $ and then sent to the prover $ \mathcal P $ and the firmware update process is initiated by the prover. Hence the opportunity to mount a helper data attack is limited to the few occasions during which a firmware update is required by the prover. In a reverse fuzzy extractor attack, the adversary is forced to perform a trial-and-error measure by continually sending malicious helper data to the resource-rich prover $\mathcal P$. The prover can perceive such malicious behavior and stop responding to a firmware update session that is hijacked by an adversary.
\item There exists a built-in throttling and obfuscation mechanism that prevents rapid helper data submissions needed for such an attack. Recall that in a man-in-the-middle attack where the adversary is able to submit manipulated helper data, the success or failure is masked by failure from external factors such as powering and the determination of success or failure is only evident at the near conclusion of the update session---this can be 2-4 seconds.
\item Under tampered helper data queries, the failure rate of reconstructing the key $\bf sk $ will significantly increase. Thus, the prover $\mathcal P$ can detect such an abnormal failure rate as a potential attack~\cite{gao2018lightweight}. This can deem the target CRFID as being compromised. Alternatively, detection can also be achieved by extending the protocol to allow a request to the token $ \mathcal T $ to send back \textsf{MAC}($nonce\|{\bf h_i}$, ${\bf sk_i}$) that the prover $ \mathcal P $ can subsequently compare using its computed $ {\bf sk_i} $ from the securely stored response $ {\bf r'} $ and the provided helper data $ {\bf h_i} $.
\end{enumerate}
{\bf Modeling attack.} We are also aware that modeling attacks on the reverse fuzzy extractors have been shown in~\cite{becker2015pitfalls,delvaux2015survey}. These attacks are applicable to arbiter PUFs (APUFs)~\cite{suh2007physical}, mainly due to the fact that the responses from APUFs are correlated. However, modeling attacks cannot be mounted on SRAM PUFs~\cite{aysu2015end} because PUF responses are information-theoretically independent since each response is derived from a spatially separate SRAM cell. 


\section{Related Work and Discussion}\label{sec:relatedwork}

\begin{table}[!b]
\centering
\textcolor{T}{\caption{Comparison Between Related Works}}
\label{tab:protocol_compare}
    \resizebox{\linewidth}{!}{
    \begin{tabular}{l| C{.6in} C{.7in} C{0.5in} C{0.6in} c c c}
    \toprule[1.5pt]
    Protocol   & Passively powered & In-application behaviour modification & Wireless firmware update  & Broadcast to multi-CRFID & Security\\ \hline
    Bootie \cite{ransford2010rudimentary}     & \ding{56}             & \ding{56}                                    & \ding{56}                  & \ding{56}       & \ding{56} \\
    FirmSwitch \cite{yang2015wireless} & \ding{52}              & \ding{52}                                   & \ding{56}                  & \ding{56}       & \ding{56} \\
    R2/R3 \cite{wu2016r2}      & \ding{52}              & \ding{52}                                   & \ding{52}                 & \ding{56}       & \ding{56} \\
    Wisent \cite{tan2016wisent}     & \ding{52}              & \ding{52}                                   & \ding{52}                 & \ding{56}       & \ding{56} \\
    Stork \cite{aantjes2017fast}      & \ding{52}              & \ding{52}                                   & \ding{52}                 & \ding{52}       & \ding{56} \\
    MSPboot \cite{ryanbrownkatiepier2016}    & \ding{56}               & \ding{52}                                   & \ding{52}                 & \ding{56}       & \ding{56} \\
    \textbf{SecuCode}   & \ding{52}              & \ding{52}                                   & \ding{52}                 & \ding{56}      & \ding{52}\\ \bottomrule[1.5pt]              
    \end{tabular}
}
\end{table}

For emerging battery-free computing devices such as CRFID platforms including WISP\cite{sample2008design}, MOO\cite{zhang2011moo} and Farsens Pyros\cite{farsens2016}, firmware updates are usually done with a wired programming interface; for example by way of a JTAG \cite{zhang2011moo} interface or a Serial interface~\cite{prasad2013wisp}. The main difficulties that hinder CRFID platform to be reprogrammed using a wireless method are: i) the transiently powered nature where encountering power failures are highly likely\cite{tan2016wisent}; ii) reprogramming code memory such as FLASH that require the device attaining an adequate voltage level from harvested power\cite{wu2016r2}; and ii) the lack of supervisory control of an operating system for managing a devices' tasks~\cite{ransford2010rudimentary}. We have seen recent efforts to bring wireless reprogramming to CRFID transponders\cite{ransford2010rudimentary,yang2015wireless,wu2016r2,tan2016wisent},  
however, to the best of our knowledge, SecuCode is the \textbf{first} work to resolve the requirement for security for wireless code dissemination for intermittently powered passive CRFID devices. Therefore, in the following we review studies to: i) develop a bootloader and modify application behavior; and ii) progress towards developing on-the fly wireless firmware update methods. \textcolor{T}{We summarize the key characteristics of these studies in~\autoref{tab:protocol_compare} and provide benchmark results for SecuCode with the \textit{non-secure} wireless update method, Wisent~\cite{tan2016wisent}, in  Appendix~\ref{Appendix:WisentComp}.}

\subsection{In-application behavior modification}

An early version of a bootloader for a CRFID platform, \textit{Bootie} was  proposed by Ransford in \cite{ransford2010rudimentary}. Bootie was designed to accept two (or more) firmware and cross-compile them into one executable to be preloaded onto a CRFID transponder. The compiled firmware was then downloaded and tested on Olimex MSP430-H2131 minimum system board. The author showed that Bootie could be used as a basis for wireless firmware updates. However, as a proof-of-concept, Bootie only enables the platform to execute pre-loaded firmwares one-by-one and did not allow responding to user demands nor operating conditions to determine the switching between firmware.

The \textit{FirmSwitch} scheme was later demonstrated in~\cite{yang2015wireless} to offer firmware flexibility for CRFID transponders. This approach allowed a user to switch between \textit{pre-loaded} firmware instances on a CRFID platform using downstream commands to the CRFID transponder. 
However, FirmSwitch was not developed to support wireless firmware updates. 


\subsection{Wireless code dissemination}



More recently, the \textit{Wisent} method by Tan {\it et al.}~\cite{tan2016wisent}, \textit{$R^2$} and \textit{$R^3$} method by Wu {\it et al.}~\cite{wu2016r2,wu2017r} demonstrated a robust wireless firmware update method for CRFID transponders using WISP platforms. In particular, $R^3$ was implemented on three different types of CRFID transponders. Subsequently, in \textit{Stork}, Aantjes {\it et al.}~\cite{aantjes2017fast} proposed a fast Multi-CRFID wireless firmware transfer protocol that involves ignoring the RN16 handle sent from an RFID transponder (i.e. the tags still save the downstream data even if their handle does not match the one specified by the reader). Stork enables an RFID Reader to simultaneously program multiple CRFID devices in the field to reduce the time to update multiple devices. Although these works achieved on the fly wireless update of firmware along with a bootloader design, none of wireless firmware update approaches address the issue of security and the trustworthiness of the prover, therefore, malicious firmware injection remains an open issue. 

In late 2016, Brown and Pier from Texas Instrument (TI) presented an application port\cite{ryanbrownkatiepier2016} extending TI's previous work, \textit{MSPBoot}\cite{luisreynoso2013}. In this work, wireless updating was demonstrated in two examples; using UART or SPI bus to interconnect an MSP430 16-bit RISC microcontroller and a CC1101 sub-1GHz RF transceiver. The enhanced bootloader design supported: i) application validation; ii) redirecting interrupt vectors; and iii) code sharing via preconfigured callbacks. Additionally, a dual image failsafe mechanism is introduced; here, before the application in executable area is overwritten, the new image would be verified in a download area. Therefore any interruption in communication would not affect the function of the device. However, like other bootloaders and wireless firmware update methods, security is not considered.

\section{Conclusion}\label{sec:conclusion}
We have presented the first secure wireless firmware update scheme, SecuCode, for resource-constrained and intermittently powered CRFID devices. We derived a volatile secret key on demand and discard it after usage to remove the difficulty of permanent secure key storage in NVM. Our SecuCode protocol only allows an authorized party to perform a wireless firmware update and does not require any hardware modifications whilst being standards compliant. As noted in~\cite{aysu2015end}, cryptographic engineering of a protocol must consider the complex environment for physical devices, such as noise and energy constraints, performance and cost of protocol instantiations. To this end, we have successfully addressed security and implementation challenges, realized an end-to-end SecuCode implementation on the popular CRFID transponder and extensively evaluated the cost and performance of our realization, including an application case study along with a complete public release of code and experimental data. 

\vspace{0.1cm}
\noindent\textcolor{T}{\textbf{Limitations:~} Our study is not without limitations.} \textcolor{T}{Since our main focus is to develop a secure end-to-end solution for wireless updates, we have only investigated an intermittent execution model based on preventing power loss where we used fixed intermittent operating settings (low power deep sleep duration) for the \textsf{FE.Gen} and \textsf{MAC} functions. These settings are determined when the bootloader is provisioned and can, unnecessarily, slow down firmware updates at short operating distances.}

\textcolor{T}{While we have addressed the problem of malicious code injection attacks, the firmware is sent as plaintext and privacy or intellectual property protection goals are not addressed by SecuCode.}
\textcolor{T}{Although Stork~\cite{parks2016} provides a method for simultaneous \textit{non-secure} dissemination of code to multiple CRIFD devices, as a first method, our approach to securely update a large number of CRFID devices requires the dissemination of code to each single device at a time.}

\vspace{0.1cm}
\noindent\textcolor{T}{\textbf{Future Work:~}}
\textcolor{T}{Since our main focus in this paper is security, we opted for an intentionally simple IEM approach where we use a fixed intermittent operating setting for the two security building blocks. Future work will investigate dynamic task scheduling where the intermittent operating setting can be set optimally as in Dewdrop~\cite{buettner2011dewdrop} to provide performance improvements such as increasing the operational range and the rate at which firmware can be transferred to a CRFID device. 
Such a dynamic task scheduling approach can investigate exploiting the decay of SRAM~\cite{rahmati2012tardis} to gather time intervals between power loss events to determine optimal task schedules. This could potentially provide an attractive solution where SRAM intrinsic to the MCU is used to determine task schedules.} 

\textcolor{T}{Additionally, in future work, we will consider the problem of secure and simultaneous code dissemination to multiple RFID devices. While an approach that excludes the requirement for key protection may provide a simpler code dissemination problem, the use of a PUF in such a context to benefit from the inherent key protection, is a challenging problem.}

\bibliographystyle{IEEEtran}
\bibliography{upRFID}

\vspace{-1cm}
\begin{IEEEbiography}[{\includegraphics[width=1in,height=1.25in,clip]{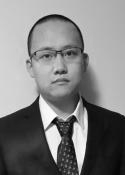}}]{Yang Su}
(S'14) Yang Su received the B.Eng. degree with first class Honours in Electrical and Electronic Engineering from The University of Adelaide, Australia in 2015. He worked as a Research Associate at The University of Adelaide from 2015-2016 and he is currently pursuing the Ph.D. degree. His research interests include hardware security, physical cryptography, embedded systems and system security. 
\end{IEEEbiography}
\vspace*{-1.4cm}
\begin{IEEEbiography}[{\includegraphics[width=1in,height=1.25in,clip]{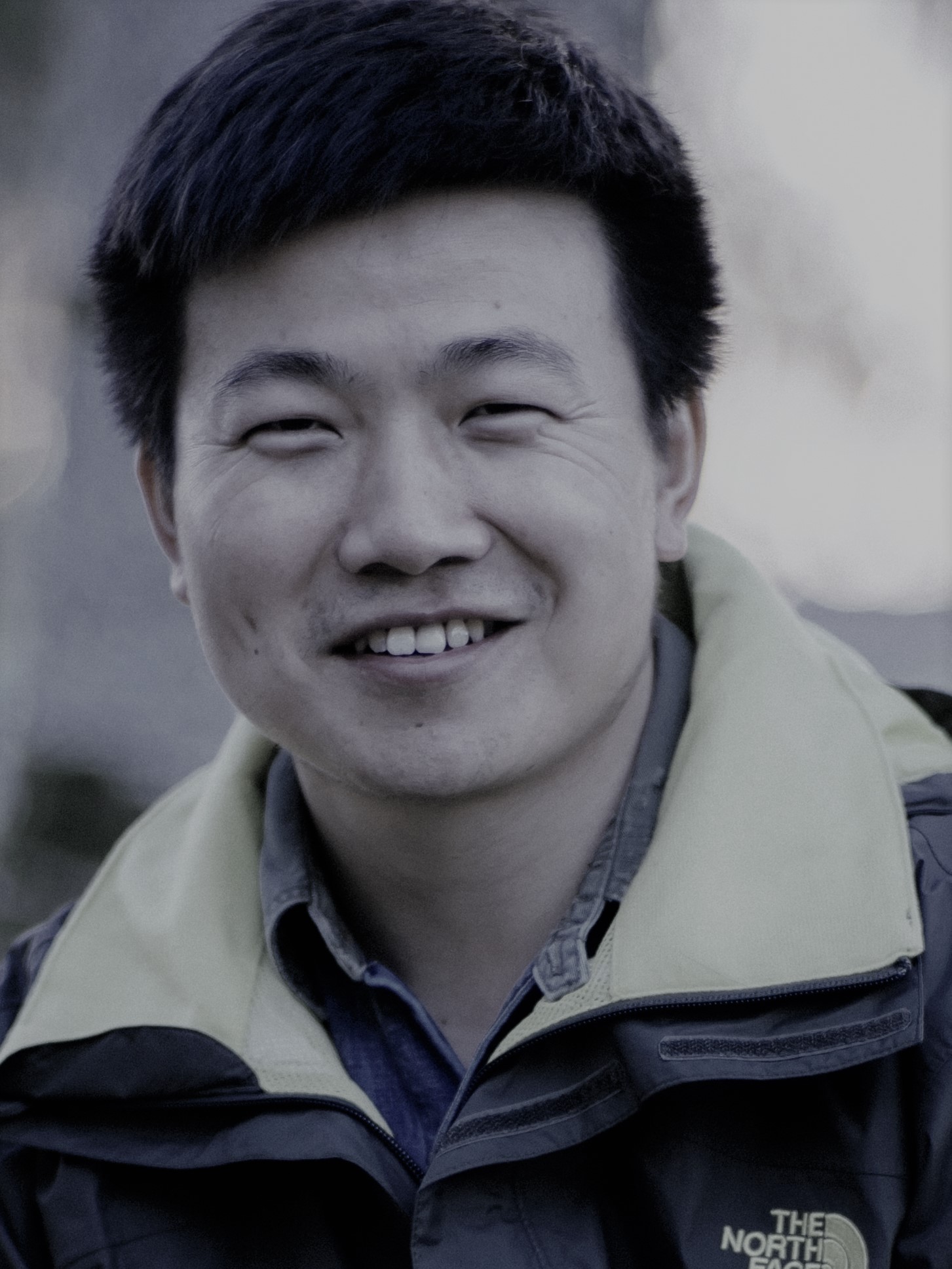}}]{Yansong Gao}
received his M.Sc degree from University of Electronic Science and Technology of China in 2013 and Ph.D degree from the School of Electrical and Electronic Engineering in the University of Adelaide, Australia, in 2017. He is now with School of Computer Science and Engineering, NanJing University of Science and Technology, China. His current research interests are hardware security and system security.
\end{IEEEbiography}

\vspace*{-1.4cm}
\begin{IEEEbiography}[{\includegraphics[width=1in,height=1.25in,clip]{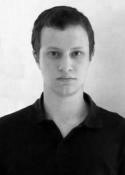}}]
{Michael Chesser} received his B.Sc Advanced degree in 2016 and his Honours (First Class) degree in Computer Science in 2017 from The University of Adelaide, Australia. Michael has worked as a consultant at Chamonix Consulting and, more recently, at the School of Computer Science, The University of Adelaide as a Research Associate. His research interests are in compilers, embedded systems, security and virtualization.
\end{IEEEbiography}

\vspace*{-1.6cm}

\begin{IEEEbiography}[{\includegraphics[width=1in,height=1.25in,clip]{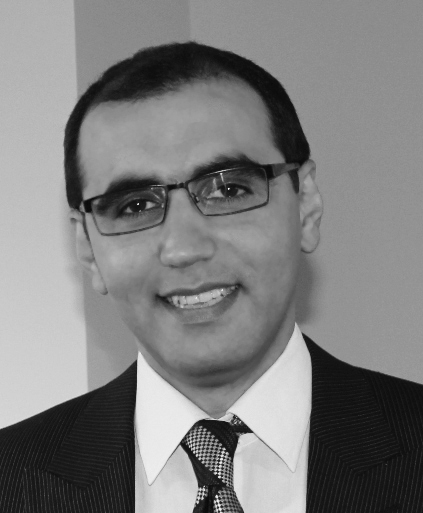}}]{Omid Kavehei}
(S'05, M'12, SM'16) received his Ph.D. degree in electronic engineering from The University of Adelaide, Australia, in 2012. He is currently a Senior Lecturer at The University of Sydney, Australia. Dr. Kavehei was an executive member of the South Australia IEEE student chapter and the recipient of The University of Adelaide’s 2013 postgraduate university alumni medal and the South Australian Young Nanotechnology Ambassador award in 2011. His research interests include emerging solid-state memory devices, physical cryptography, novel computational paradigms based on nanotechnology. He is a Senior Member of IEEE.
\end{IEEEbiography}

\vspace*{-1.3cm}

\begin{IEEEbiography}[{\includegraphics[width=1in,height=1.25in,clip]{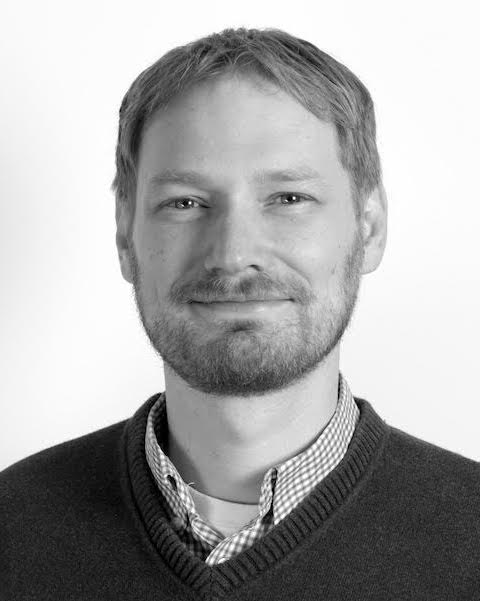}}]{Alanson Sample} received his Ph.D. in electrical engineering from the University of Washington in 2011. He is currently an Associate Professor in the computer science and engineering division of the department of Electrical Engineering and Computer Science at the University of Michigan. Prior to returning to academia, he spent the majority of his career working in academic minded industry research labs.  Most recently Alanson was the Executive Lab Director of Disney Research in Los Angeles and before that the Associate Lab Director and a Principal Research Scientist at Disney Research, Pittsburgh. Prior to joining Disney, he worked as a Research Scientist at Intel Labs in Hillsboro. He also held a postdoctoral research position in the Department of Computer Science and Engineering at the University of Washington. Throughout his academic studies he worked fulltime at Intel Research, Seattle. Dr. Sample's research interests lie broadly in the areas of Human-Computer Interaction, wireless technology, and embedded systems. 
\end{IEEEbiography}

\vspace*{-1.3cm}

\Urlmuskip=0mu plus 1mu\relax

\begin{IEEEbiography}[{\includegraphics[width=1in,height=1.25in,clip]{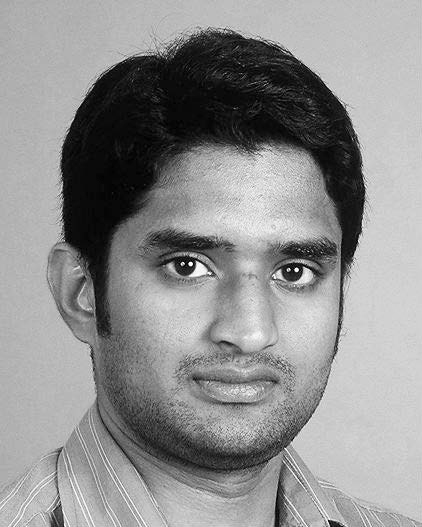}}]{Damith C. Ranasinghe}
received the Ph.D. degree in electrical and electronic engineering from The University of Adelaide, Australia, in 2007. From 2005 to 2006, he was a Visiting Scholar with the Massachusetts Institute of Technology and a Post-Doctoral Research Fellow with the University of Cambridge from 2007 to 2009. He joined The University of Adelaide in 2010, and is currently an Associate Professor with the School of Computer Science. His research interests include pervasive computing, autonomous systems, deep learning, and security.
\end{IEEEbiography}

\clearpage
\appendices
\setcounter{page}{1}




\begin{table}[!ht]
\centering
\caption{Random Number Generator Comparison} 
\label{tab:RNG}
\resizebox{\columnwidth}{!}{%
\begin{tabular}{ C{0.8in} C{.4in} C{.5in} C{.5in} C{.6in}}\toprule[1.5pt]
\bf Name & \bf Bits per request & \bf Time overhead & \bf Energy consumption  &\bf Hardware requirement\\\midrule
CTR-DRBG        &  128     & 214 us      &  14.83 nJ     	& AES ,FRAM \\
Thermal noise	&  16	& 27.2 us		& 2.69 nJ			& ADC\\
SRAM-TRNG		&  128	& 23.10 us	&2.32 nJ	& --	\\
\bottomrule[1.25pt]\\
\end {tabular}
}

    \raggedright 
    Note: the calculations are based on the technical details obtained from MSP430FR5969 datasheet for a 1~MHz MCLK and 3.3~V power supply.
\end{table}

\vspace{-0.5cm}
\section{Random Number Generator Comparison}
\label{Sec:APX-TRNG}

We provide a summary of the random number generator methods we have evaluated.
\newline
\newline
{\bf CTR-DRBG:} Texas Instrument has demonstrated a Counter Mode Deterministic Random Byte Generator (CTR-DRBG) in~\cite{akashpatelcaleboverbay2017} to meet the needs of security mechanisms on MSP430 MCUs. 
The CTR-DRBG is developed by National Institute of Standard and Technology (NIST) in NIST SP 800-90A~\cite{barker2012nist} and is built upon a block cipher algorithm (e.g., AES256). The CTR-DRBG passes all NIST randomness test criteria; implying that the CTR-DRBG has comparable performance to a true random number generator. \\
{\bf Thermal noise:} Thermal noise (Johnson–Nyquist noise) has been exploited \cite{callegari2005embeddable}\cite{lampert2016robust}\cite{petrie2000noise} as an entropy source for a TRNG. Thermal noise results in unpredictable small voltage fluctuations in resistive components at any temperature above 0°K. The least significant bit (LSB) of sampled data is significantly affected by  thermal noise over other bits\cite{shen2010practical}\cite{bucci2003high}. Therefore, random bits can be extracted by sampling the LSB from a noisy sensor. Random number generators from this method has passed the NIST randomness tests~\cite{yan2016cryptographic}.\\
{\bf Noisy SRAM responses:} The noisy SRAM PUF responses can be used to generate true random numbers; this method has been extensively studied, for example in~\cite{aysu2015end,holcomb2009power}. Sufficient entropy can be extracted through, e.g., XOR and bit shift operations\cite{aysu2015end,holcomb2009power} over a number of SRAM PUF responses. Random bit streams from this method has passed the NIST random number generator test suite~\cite{holcomb2009power}. 


\begin{table}[t]
\centering
\caption{CRP-Block efficiency} 
\label{tab:MBeffi} 
\begin{tabular}{ C{1in} C{.5in} C{.5in} C{.5in}}\toprule[1.5pt]
\bf Chip ID & \bf Blocks produced & \bf Available PUF bits & \bf Efficiency\\\midrule
1 & 4 & 992 & 11.2\%\\
2 & 8 & 1984 & 22.3\%\\
3 & 2 & 496 & 5.58\%\\
4 & 2 & 496 & 5.58\%\\
5 & 8 & 1984 & 22.3\% \\
6 & 7 & 1736 & 19.5\% \\
7 & 2 & 496 & 5.58\%\\
8 & 5 & 1240 & 13.9\%\\
9 & 3 & 744 & 8.36\%\\
10 & 6 & 1488 & 16.7\%\\
11 & 6 & 1488 & 16.7\%\\
12 & 6 & 1488 & 16.7\%\\
13 & 6 & 1488 & 16.7\%\\
14 & 7 & 1736 & 19.5\% \\
15 & 6 & 1488 & 16.7\%\\
16 & 8 & 1984 & 22.3\% \\
17 & 7 & 1736 & 19.5\% \\
18 & 4 & 992 & 11.2\%\\
19 & 6 & 1488 & 16.7\%\\
20 & 6 & 1488 & 16.7\%\\
WISP5.1 LRG & 7 & 1736 & 19.5\% \\

\bottomrule[1.25pt]
\end {tabular}\par
\end{table}

\begin{table*}[!ht]
\centering
\caption{MAC evaluation with LED firmware size = 153 bytes.}
\label{tab:hashShort}
\resizebox{\textwidth}{!}{%
\begin{tabular}{l|llllll}\toprule[1.5pt]
MAC      & Digest size (bits) & Byte in digest & Clock Cycles & Cycle per message byte & Code size (bytes) & Internal state size (bytes) \\\hline
BLAKE2s-256	& 256	&32	&81,036	&530	&4,964	&238\\
BLAKE2s-128 & 128	&16	&79,276 	&518	&4,961 &238\\
HWAES-GMAC	&128	&16	&432,337 & 2,826	&3,538	&268\\
HWAES-CMAC	&128	&16	&15,876	& 104	&3,198	&58\\
\bottomrule[1.25pt]                                    
\end{tabular}
}
\end{table*}


\begin{table*}[!ht]
\centering
\caption{MAC evaluation of Accelerometer firmware size = 419 bytes.}
\label{tab:hashLong}
\begin{tabular}{l|llllll}\toprule[1.5pt]
MAC      & Digest size (bits) & Byte in digest & Clock Cycles & Cycle per message byte & Code size (bytes) & Internal state size (bytes) \\\hline
BLAKE2s-256	& 256	&32	&183,230	& 437	&4,964	&238\\
BLAKE2s-128 & 128	&16	&181,470 	& 433	&4,961 &238\\
HWAES-GMAC	&128	&16	&1,106,538	& 2,641	&3,538 &268\\
HWAES-CMAC	&128	&16	&37,041	& 88	&3,198 &58\\
\bottomrule[1.25pt]     

\end{tabular}
\end{table*}

\section{SRAM PUF Pre-selection Efficiency}\label{app:selectioneff}
Efficiency evaluates the ratio of number of the selected reliable bits in the CRP Blocks selected over the total number of possible response bits (8,896 bits; obtained by deducting 2,048 bits reserved for the stack, 5,440 bits for the SRAM-TRNG and static variables from 16,384 SRAM bits). We tested 20 new MSP430FR5969 chips and one WISP5.1LRG. The result are summarized in Table~\ref{tab:MBeffi}. The efficiency depends on the noise level of the SRAM cells in the candidate chip, if a chip cannot provide adequate number of CRP Blocks, it should be excluded before deployment. Nonetheless, for 20 tested chips, at least two independent CRP-blocks---each consisting of 248 bits---are obtained. 

\section{MAC function evaluations}\label{Appendix:mac}
\textcolor{T}{Notably, benchmarks of software implementations of MAC function on desktop platforms are not suitable since these implementations use specific CPU instructions, such as the SSE instruction set\cite{lemire2016faster}, and advanced paradigms like out-of-order execution, which are not supported on resource-constrained embedded systems. Therefore, for the first time, we benchmarked a set of MAC functions on MSP430FR5969. Besides HWAES-GMAC, we also considered BLAKE2s as competitor.} 

\textcolor{T}{The MAC function tests were based on two WISP firmwares; i) LED firmware (short string); and ii) 3-axis accelerometer firmware (long string). Table~\ref{tab:hashShort}, and \ref{tab:hashLong} detail our results. We selected HWAES-CMAC to obtain a 128 bit MAC in our SecuCode implementation.} 

\begin{figure*}[t]  
\centering
\includegraphics[width=0.82\textwidth]{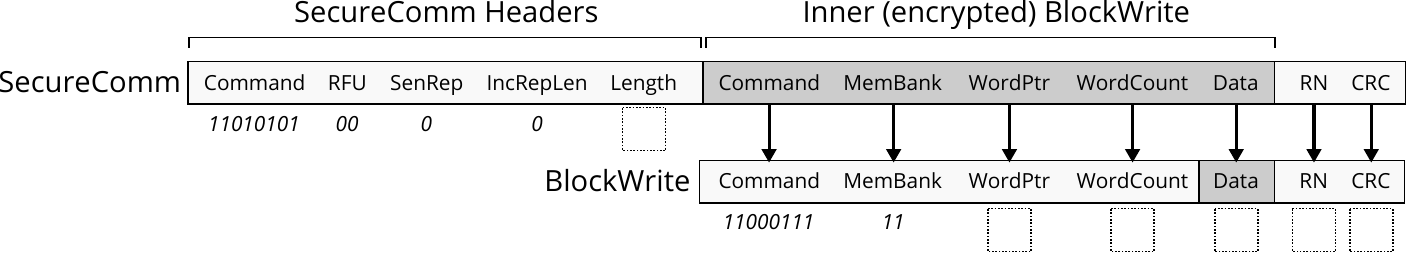}
\caption{\texttt{SecureComm} command encoding. \texttt{SecureComm} specification is defined to encapsulate an encrypted command. Instead of encrypting the entire inner \texttt{BlockWrite} command, we encrypt the data field and restrict the \texttt{WordPtr} to only allow writing to the Download Area.}
\label{fig:securecomm-mapping}
\end{figure*}

\begin{figure*}[t] 
\centering
\includegraphics[width=0.79\textwidth]{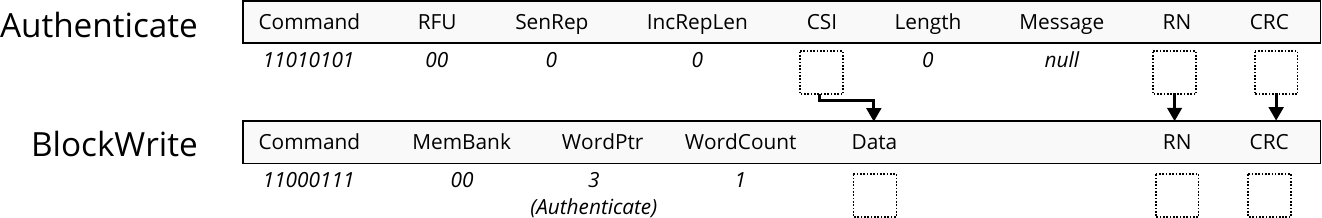}
\caption{\texttt{Authenticate} command encoding. We use \texttt{MemBank} = 0 (Reserved) in the \texttt{BlockWrite} command to indicate that the message is not a regular \texttt{BlockWrite} and \texttt{WordPtr} = 3 to indicate that the command should be processed as an \texttt{Authenticate} command. The 8-bit \texttt{CSI} field is placed in the 16-bit \texttt{Data} field of the \texttt{BlockWrite} command, and the \texttt{RN} and \texttt{CRC} fields are as specified for the \texttt{BlockWrite} command.}
\label{fig:authenticate-mapping} 
\end{figure*}

\section{EPC Global C1G2v2 command encoding}
\label{sec:commandCoding}
The Impinj R420 RFID reader used in our experiments does not yet support the recent \epc protocol changes, hence,  we map the unsupported commands to \texttt{BlockWrite} commands. The mapping for \texttt{Authenticate} and \texttt{SecureComm} commands are detailed in Fig.~\ref{fig:authenticate-mapping} and Fig.~\ref{fig:securecomm-mapping}, respectively. Similar to \texttt{Authenticate}, a \texttt{TagPrivilege} command is mapped to a \texttt{BlockWrite} command with \textit{WordPtr}~=~0x7E. Additionally, the Impinj reader does not have complete support for a  \texttt{BlockWrite} command as outlined in the \epc protocol specification. Internally, a \texttt{BlockWrite} command containing more than 1 word (16 bits) is split into multiple smaller \texttt{BlockWrite} commands; each of which needs to be \textit{ACK}ed by the CRFID transponder. This is handled in the RFID MAC layer on a CRFID device. 
In our experiments, we still send multi-byte \texttt{BlockWrite} commands to the reader from the host running the SecuCode App as it reduces the reader-to-host communication overhead by allowing the reader to send the next \texttt{BlockWrite} as soon as an \textit{ACK} is received.

\section{Experimental Methods}\label{Appendix:expSetup}

We devised a simple method to evaluate the success rate. 
We used the ratio of the number of occurrences of backscatter events immediately after a cold start for a fixed number of cold-start attempts as the \textit{success rate}. Each cold start attempt was realized with an RFID reader interrogation signal; i.e. conducting an inventory round using an RFID reader. 
In order to mitigate the possible network delays in using a networked RFID reader connected to host machine to determine the delays, we employed two oscilloscope probes directly connected to 
the CRFID device to measure reservoir capacitor voltage and backscatter events to determine the time at which the backscatter event is initiated as well as to determine if the backscatter event immediately followed a cold start. 


In order to independently evaluate key derivation and hash function execution, we created a second bootloader instance where the physically obfuscated key derivation code was replaced with the execution of a selected hash function. This allowed us to use the same method described to ascertain the success rate of the hash function execution under wireless powering conditions.

\begin{figure*}[t]
    \centering
    \includegraphics[width=\textwidth]{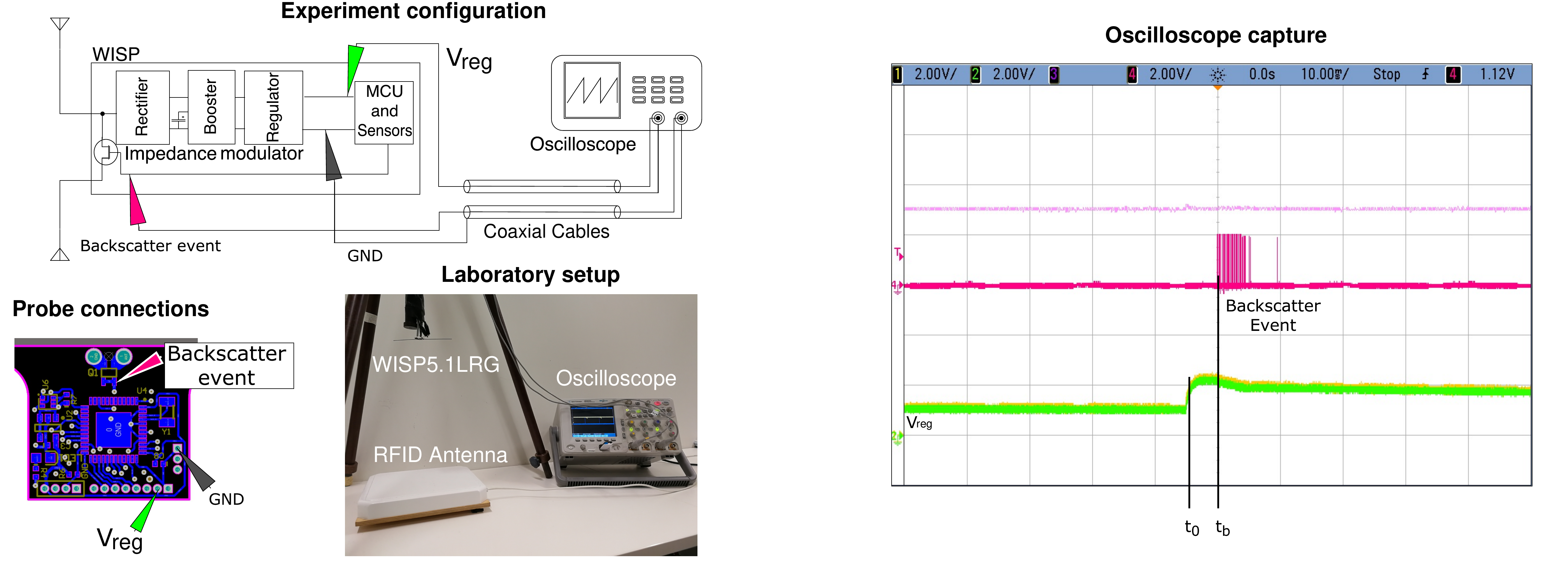}
    \caption{Experiment setup for measuring latency and success rate for PUF key derivation and MAC function executions.}
    \label{fig:expSetup}
\end{figure*}

The success rate and latency of both lightweight physically obfuscated key derivation method and hash function executions are evaluated using the experimental setup shown in Fig.\ref{fig:expSetup}. Two probes are connected to the CRFID circuit board where the common ground (GND) is colored in black in the figure. The backscatter signal probe lead is colored in red, the regulated voltage ($V_{\rm reg}$) is colored in green. We define the time latency as the interval between the $V_{\rm reg}$ reaching 2.0~V ($t_0$) and the backscatter event ($t_b$). An operation is successful if a backscattering event occurs before $V_{\rm reg}$ drops below 1.8~V---minimum operating voltage of the MCU.

\begin{figure*}[t]
    \centering
    \includegraphics[width=.75\linewidth]{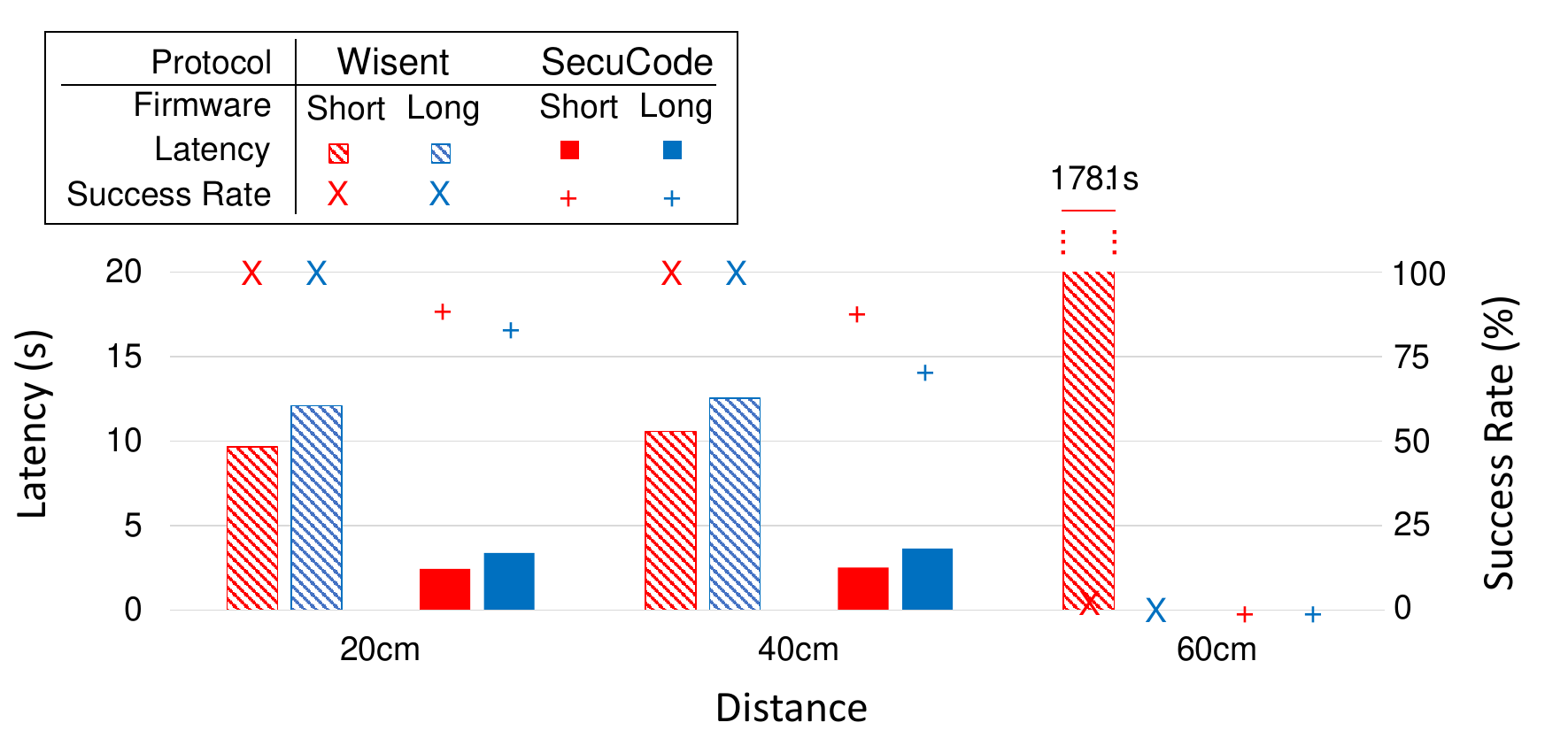}
  \textcolor{T}{\caption{Comparing latency and success rate of SecuCode and Wisent \cite{tan2016wisent}.}}
  \label{fig:vsWisentSuccessRateLatency}
  \includegraphics[width=.75\linewidth]{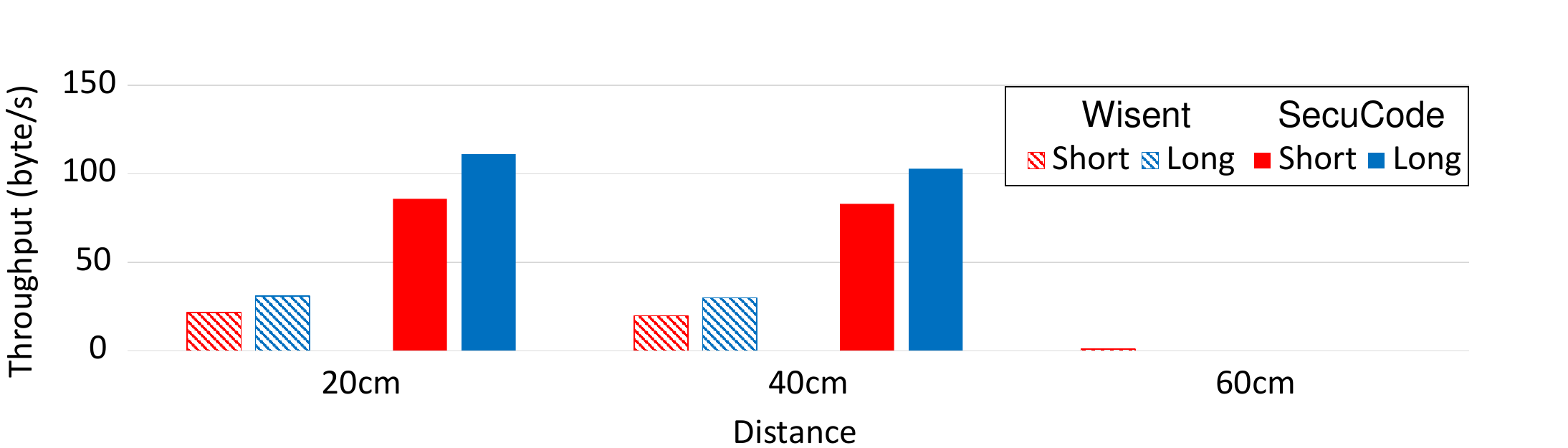}
  \textcolor{T}{\caption{Comparing the throughput of SecuCode and Wisent \cite{tan2016wisent}.}}
  \label{fig:vsWisent}
\end{figure*}

\textcolor{T}{\section{End-to-end Comparison}\label{Appendix:WisentComp}}
\textcolor{T}{We selected Wisent~\cite{tan2016wisent} as the non-secure protocol to benchmark SecuCode against, because: i) Wisent focuses on the dissemination of firmware, albeit non-secure, to a \textit{single} device, as we do; ii) both projects employ the the same CRID device; and iii) Wisent code is publicly available. Notably, a comparison of Wisent to Stork~\cite{parks2016}, capable of broadcasting code to many CRFID devices, is found in~\cite{parks2016} already.}

\textcolor{T}{It is difficult to make a direct comparison since the secure and non-secure approaches have fundamentally different goals and tradeoffs. For example: i) the write and check method in Wisent vs. write all and validate method of SecuCode; and ii) use of a custom bootloader in Wisent vs. our bootloader design built on the MSPboot framework of Texas Instruments for industry compliance. 
However, to provide an understanding of the overhead the secure method, we provide measurements for three performance measures: i) mean latency to successfully transfer a given firmware; ii) success rate; and iii) throughput (successful firmware bytes written/time taken).}

\textcolor{T}{Both Wisent and SecuCode require a pre-installed bootloader which includes the WISP 5 base firmware (5600 to 5700 bytes). The Wisent bootloader requires an additional 608 bytes of code memory, while the SecuCode bootloader requires 6024 bytes of code memory as a consequence of additional security routines\footnote{These values are dependent on optimization settings/compiler versions and are only approximate.}. 
In Fig.~\ref{fig:vsWisentSuccessRateLatency} and in Fig.~\ref{fig:vsWisent}, we compare the end-to-end performance at three different operating distances. These results are obtained from evaluations based on 100 repeated firmware update attempts. We used two firmware sizes: i) Short (210 bytes); and ii) Long (376 bytes).} 

\textcolor{T}{SecuCode and Wisent configure the reader to transmit the same \texttt{BlockWrite} multiple times to increase \texttt{BlockWrite} success rate without the overhead of checking the result on the host. In addition Wisent includes a checksum for each block, if the checksum does not match then Wisent re-sends the block. Notably, SecuCode only checks that the \texttt{BlockWrite} command was ACKed by the tag.}

\textcolor{T}{Wisent is able achieve 100\% success rate at 20 cm and 40 cm, however the overhead of the per-block checksum and message header decreases the throughput. SecuCode only validates data integrity once the firmware has been completely transmitted; and refusing the firmware update if the integrity check fails. Further, any power loss event causes SecuCode to enter into a new firmware update sessions. Consequently, the success rate for SecuCode varies between 73\% and 89\% and depends on firmware size. However, compared to Wisent, SecuCode has better throughput. Notably, Wisent attained only one successful firmware update out of 100 trails at 60 cm and the update took 178.1 seconds to complete with a resulting throughput 1.18 byte per second.  While SecuCode ceased to complete the firmware update successfully at 60 cm due to brownout.}

\end{document}